\documentstyle[11pt,epsfig,aaspp4]{article}
\tighten 
\begin{document}

\baselineskip 5.0ex

\newcommand{\lya}{Lyman~$\alpha$}
\newcommand{\lyb}{Lyman~$\beta$}
\newcommand{\za}{$z_{\rm abs}$}
\newcommand{\ze}{$z_{\rm em}$}
\newcommand{\cmtwo}{cm$^{-2}$}
\newcommand{\nhi}{$N$(H$^0$)}
\newcommand{\nzn}{$N$(Zn$^+$)}
\newcommand{\ncr}{$N$(Cr$^+$)}
\newcommand{\degpoint}{\mbox{$^\circ\mskip-7.0mu.\,$}}
\newcommand{\halpha}{\mbox{H$\alpha$}}
\newcommand{\hbeta}{\mbox{H$\beta$}}
\newcommand{\hgamma}{\mbox{H$\gamma$}}
\newcommand{\kms}{\,km~s$^{-1}$}      % note leading thinspace
\newcommand{\minpoint}{\mbox{$'\mskip-4.7mu.\mskip0.8mu$}}
\newcommand{\mv}{\mbox{$m_{_V}$}}
\newcommand{\Mv}{\mbox{$M_{_V}$}}
\newcommand{\peryr}{\mbox{$\>\rm yr^{-1}$}}
\newcommand{\secpoint}{\mbox{$''\mskip-7.6mu.\,$}}
\newcommand{\sqdeg}{\mbox{${\rm deg}^2$}}
\newcommand{\squig}{\sim\!\!}
\newcommand{\subsun}{\mbox{$_{\twelvesy\odot}$}}
\newcommand{\et}{et al.~}

\def\ltsima{$\; \buildrel < \over \sim \;$}
\def\simlt{\lower.5ex\hbox{\ltsima}}
\def\gtsima{$\; \buildrel > \over \sim \;$}
\def\simgt{\lower.5ex\hbox{\gtsima}}
\def\arcs{$''~$}
\def\arcm{$'~$}
\vspace*{0.1cm}
\title{THE METALLICITY OF HIGH REDSHIFT GALAXIES:\\
THE ABUNDANCE OF ZINC IN 34 DAMPED LYMAN $\alpha$ SYSTEMS
FROM $z$ = 0.7 TO 3.4}

\vspace{1cm}
\author{\sc Max Pettini}
\affil{Royal Greenwich Observatory, Madingley Road, Cambridge, CB3 0EZ, UK}
\affil{e-mail: pettini@ast.cam.ac.uk}
\author{\sc Linda J. Smith}
\affil{Department of Physics and Astronomy, University College
London,}
\affil{Gower Street, London WC1E 6BT, UK}
\affil{e-mail: ljs@star.ucl.ac.uk}
\author{\sc David L. King}
\affil{Royal Greenwich Observatory, Madingley Road, Cambridge, CB3 0EZ, UK}
\affil{e-mail: king@ast.cam.ac.uk}
\author{\sc Richard W. Hunstead}
\affil{School of Physics, University of Sydney, NSW 2006, Australia}
\affil{e-mail: rwh@physics.usyd.edu.au}

\newpage
\begin{abstract}
We report new observations of Zn~II and Cr~II 
absorption lines in 10 damped \lya\ 
systems (DLAs), mostly at redshift $z_{\rm abs} \simgt 2.5$\,. 
By combining these results with those from our earlier survey 
(Pettini et al. 1994) and other recent data, 
we construct a sample of 34 
measurements (or upper limits) of the Zn abundance relative to hydrogen
[Zn/H]; 
the sample includes more than one 
third of the total number of DLAs known.

The plot of the abundance of Zn as a function of redshift 
reinforces the two main  
findings of our previous study. 
(1) Damped \lya\ systems are 
mostly metal-poor, {\it at all redshifts sampled}; the column density 
weighted mean for the whole data set is 
[Zn/H] $= -1.13 \pm 0.38$ (on a logarithmic scale),
or approximately 1/13 of solar. 
(2) There is a large spread, by up to two orders of 
magnitude, in the metallicities we measure
at essentially the same redshifts. 
We propose that damped \lya\ systems are drawn from a varied population
of galaxies of different morphological types and at different 
stages of chemical evolution, supporting the idea of a protracted epoch 
of galaxy formation.

At redshifts $z \simgt 2$ the typical metallicity of 
the damped \lya\ systems
is in agreement with expectations based
on the consumption of H~I gas implied by the recent 
measurements of $\Omega_{\rm DLA}$ by Storrie-Lombardi et al. (1996a),
and with the metal 
ejection rates in the universe at these epochs 
deduced by Madau (1996) from the ultraviolet luminosities 
of high redshift galaxies revealed by deep imaging surveys. 
There are indications in our data
for an increase in the 
mean metallicity of the damped \lya\ systems from 
$z >  3$ to $\approx 2$, consistent with
the rise in the comoving star formation rate
indicated by the relative numbers of $U$
and $B$ drop-outs in the {\it Hubble Deep Field}.
Although such comparisons are still tentative, 
it appears that these different avenues for
exploring the early evolution of galaxies give a broadly consistent 
picture. 

At redshifts $z < 1.5$ DLAs evidently do not exhibit the 
higher abundances expected from a simple closed-box 
model of global chemical evolution, 
although the number of measurements is still very small. 
We speculate that this may be due to an increasing contribution of low 
surface brightness galaxies to the cross-section for 
damped \lya\ absorption and to the increasing dust bias with decreasing 
redshift proposed by Fall and collaborators. 
However, more DLAs at 
intermediate redshifts need to be 
identified before the importance of these effects can be assessed 
quantitatively.

The present sample is sufficiently large for a first attempt at
constructing the metallicity distribution of damped \lya\ systems
and comparing it with those of different stellar populations of the Milky Way.
The DLA abundance histogram is both broader and peaks at 
lower metallicities that those of either thin or thick disk
stars.  At the time when our Galaxy's metal enrichment was at levels 
typical of DLAs, its kinematics were closer to those of the 
halo and bulge than a rotationally supported disk.
This finding is at odds with the 
proposal that most DLAs are large disks with rotation velocities in excess of 
200~\kms, based on the asymmetric profiles of absorption lines 
recorded at high spectral resolution.
Observations of the familiar optical emission lines from H~II regions, 
which are within reach of near-infrared spectrographs on 8-10~m 
telescopes, may help resolve this discrepancy.\\
\end{abstract}

\newpage
\section{INTRODUCTION}

In the last twelve months there has been a dramatic increase in our 
ability to identify normal
galaxies at $z \simeq 3$, study their stellar populations,
and measure the rates of star formation and metal production 
in the universe over most of the 
Hubble time (Steidel et al. 1996; Madau et al. 1996).
The most prominent features in the spectra of field galaxies
at high-redshift (as is the case in the ultraviolet spectra of nearby 
star-forming galaxies) are strong interstellar lines which are  
similar, both qualitatively (in the range of ionization stages seen) and 
quantitatively (in the strengths of the absorption), to those 
in damped \lya\ systems; this similarity is consistent with the view 
that this class of QSO absorbers traces the material available 
for star formation at $z \simgt 2$  (e.g. Wolfe 1995).
The connection between normal galaxies and damped \lya\ systems
(DLAs) is a particularly important one to make and 
clarifying several aspects of this connection remains a 
priority. The reason is simple:
QSOs with known DLAs are typically 
more than 5 magnitudes brighter than a
$L^{\ast}$ galaxy at the same redshift. 
Consequently, we will inevitably continue to 
rely mostly on QSO absorption line spectroscopy for the study of physical 
conditions in the early stages of galaxy formation.

Since 1990 (Pettini, Boksenberg, \& Hunstead 1990) we have been 
conducting a survey of metallicity and dust in DLAs taking advantage of 
the diagnostic value of weak transitions of Zn~II and Cr~II.
As explained in that paper (see also the critical reappraisal 
of the technique in Pettini et al. 1997), 
[Zn/H]\footnote{We use the conventional notation where [X/Y] =
log~(X/Y)$-$log~(X/Y)$_{\sun}$} is a straightforward measure of the degree 
of metal enrichment analogous to the stellar [Fe/H], while 
[Cr/Zn] reflects the extent to which grain constituents are removed from 
the gas phase and thereby gives an indication of the
dust-to-metals ratio. 
The major results of the survey were reported in Pettini et al. (1994).
From the analysis of Zn and Cr abundances in 17 DLAs, mostly at $z \simeq 
2$, we concluded that the typical metallicity of the universe 
at a look-back time of $\sim  13$~Gyr 
($H_0 = 50$~\kms\  Mpc$^{-1}$;  $q_0  =  0.01$)
was $Z_{\rm DLA}  =  1/10  Z_{\sun}$. 
We further found that there is 
a considerable range---by up to two orders of magnitude---in  
the  degree of metal enrichment reached by different
damped \lya\ galaxies at essentially the same epoch,
and that even at 
these early stages of galaxy formation 
dust appears to be an important component of the interstellar medium, 
leading to the selective depletion of refractory elements from the gas. 

A natural next step is to extend the Zn and Cr abundance measurements 
over a wider range of redshifts than that considered by Pettini et al. (1994)
with the ultimate aim of identifying the emergence of heavy elements 
and dust in
galaxies and following their build-up with time.
To this end we have continued our 
survey since 1994; the full sample now consists of 34 DLAs, 
more than one third of the total number known (Wolfe et al. 1995).
In this paper we present the new data and consider the conclusions that 
can drawn from 
the whole set of measurements of [Zn/H]; preliminary reports have 
appeared in conference proceedings (e.g. Pettini et al. 1995a;
Smith et al. 1996). 
Our findings on the abundance of dust from consideration of the 
[Cr/Zn] ratio in the same sample have been reported separately
(Pettini et al. 1997).
Recently, Lu et al. (1996) have addressed similar questions 
from measurements of [Fe/H] in 20 DLAs using high-resolution 
echelle spectra acquired with the Keck
telescope. 
These authors reach 
conclusions which are in agreement with those presented here regarding the 
emergence of heavy elements at high redshifts, although the analysis of  
[Fe/H] is complicated by the fact that this ratio, unlike [Zn/H],
depends on {\it both} the 
metallicity and dust content of the interstellar medium.

Before proceeding it is useful to point out that in cases where
DLAs from the present sample have been reobserved with HIRES on
Keck (Wolfe et al. 1994; Wolfe 1995; Prochaska \& Wolfe 1997a),
[Zn/H] has been found to be in good agreement with the values
measured in our survey, which is based on 4-m telescope data (see
\S2 below). While the exceptional quality of the Keck
observations has made possible several new aspects of this work,
including the study of the relative abundances of a wide range of
elements and the analysis of the kinematics of the absorbing gas,
the basic survey of metallicity in DLAs can be carried out
satisfactorily with 4-m class telescopes. The main reason for
this is the optically thin nature of the Zn~II and Cr~II lines in
most DLAs  proposed by Pettini et al. (1990) and confirmed by
subsequent Keck spectra.\\

\section{OBSERVATIONS AND DATA REDUCTION}

The new data reported in this paper consist of observations of 10 DLAs in 
9 QSOs obtained between March 1994 and February 1996 (an additional 
candidate DLA from the low dispersion survey by Storrie-Lombardi et al. 
(1996b)---at \za =3.259 in the \ze =4.147 BAL QSO 1144$-$073---was shown 
not to be a damped
system by our higher resolution observations of the \lya\ absorption line).
In Table 1 we have collected relevant information for the 10 DLAs; 
the references listed in column (4) are the papers where the damped nature 
of the absorber was first identified. 
The absorption redshifts measured from associated 
metal lines in our blue and red spectra are listed in column (5); with 6 
new DLAs at $z_{\rm abs} > 2.5$, we have tripled the number of absorbers 
in this redshift regime compared with our earlier sample.

The observations, reduction of the spectra and derivation of Zn and Cr 
abundances followed the procedures described in Pettini 
et al. (1994) and the interested reader is referred to that paper
for a detailed treatment.
Briefly, the observations were carried out mostly with the double-beam
cassegrain spectrograph
of the William Herschel telescope on La Palma, Canary Islands; additional 
red spectra were secured with the cassegrain spectrograph of the 
Anglo-Australian telescope at Siding Spring Observatory, Australia.
At $z_{\rm abs} > 2.5$ the Zn~II~$\lambda\lambda 2025.483, 2062.005$ and 
Cr~II~$\lambda\lambda 2055.596, 2061.575, 2065.501$ multiplets
are redshifted longwards of 7175~\AA, where the quantum efficiency of 
CCDs falls with increasing wavelength. 
Using EEV and Tektronix CCDs we generally 
found it necessary to integrate for longer than $\sim 20\,000$~s 
(column 8 of Table 1) in order to achieve S/N between 9 and 46
(column 9).
With a spectral resolution of 0.75--1.1~\AA\ FWHM (column 7),
the corresponding $3\sigma$ detection limits for the rest frame 
equivalent widths of unresolved Zn~II and Cr~II absorption lines
range from $W_0$($3\sigma$) = 66 to 14~m\AA\ (column 10).
The final ``depth'' of the survey---that is the lowest metallicity that 
can be measured---depends on the combination of $W_0$($3\sigma$)
and the neutral hydrogen column density $N$(H$^0$). Since the values of 
$N$(H$^0$) in the new DLAs observed span one order of magnitude
(see \S3 below),
it is the sight-lines with the largest column densities of gas which 
provide the most stringent limits on metal abundances.  
Accordingly, we have tended to select DLAs for the present survey
primarily on the basis of the value of $N$(H$^0$).

In Figure 1 we have reproduced 
portions of the QSO spectra encompassing the regions 
where the Zn~II and Cr~II lines are expected
in the 10 DLAs in Table 1.
As can be seen from the figure, the absorption lines sought are 
detected in approximately half of the cases. 
Table 2 lists redshifts and 
rest-frame equivalent widths for the detections;
in the other cases the $3\sigma$ limits given in 
column (10) of Table 1 apply.

With the double-beam spectrograph on the WHT we were able to record 
portions of the blue spectrum of each QSO, centred on the damped \lya\ 
line, simultaneously with the red arm observations aimed at the 
Zn~II and Cr~II lines. 
The blue detector was either the Image Photon Counting System or a thinned 
Tektronix CCD; exposure times were the same as those given in column (8) 
of Table 1.
A 600 grooves/mm grating was used to record a 800~\AA\ 
wide interval of the spectrum with a resolution of $\sim 1.5$~\AA\ FWHM.
This configuration was chosen in preference to the higher resolving power 
achievable with a 1200 grooves/mm grating because a good definition of 
the QSO continuum is a key factor in determining the accuracy 
with which $N$(H$^0$) can be deduced from the profile of the damping wings 
of the \lya\ absorption line.
Normalised portions of the blue spectra are shown in Figure 2 together 
with our fits to the damped \lya\ lines.
The theoretical damping profiles are centred at the redshifts of the
O~I~$\lambda 1302.1685$ lines which are encompassed
by our blue data.\\

\section {ZINC AND CHROMIUM ABUNDANCES}

The main results of our survey are collected in Table 3 which includes
the 10 DLAs in Table 1 and 7 additional
systems for which data have been published
since our earlier 
study (Pettini et al. 1994). 
Values of the neutral hydrogen column density $N$(H$^0$) are listed in 
column (3) of Table 4; the typical accuracy of these measurements,
including the uncertainty in the placement of the continuum,
is $\pm 20$\%.
$N$(H$^0$) is likely to account for most of the neutral 
gas in each DLA 
given the low molecular fractions which apply to these absorbers at high 
redshifts (Levshakov et al. 1992; Ge \& Bechtold et al. 1997;
\'{C}irkovi\'{c} et al. 1997).

Columns (3) and (6) of Table 3 give the column densities 
of Zn$^+$ and Cr$^+$ respectively, deduced from the measured equivalent 
widths (or upper limits) assuming no line saturation.
That this is 
generally the case is indicated by: (1) the weakness of the absorption lines; 
(2) the equivalent width ratios of lines within each multiplet which, 
when measurable, are usually close to the ratios of the corresponding 
$f$-values (Bergeson \& Lawler 1993); 
and (3) the resolved absorption profiles recorded 
with HIRES on Keck for many DLA systems, 
including some in common with the present survey
(Lu et al. 1996; Prochaska \& Wolfe 1997a).
There are of course exceptions, such as 
the $z_{\rm abs} = 2.5842$ system in Q1209$+$093---see the discussion at 
\S3.12 below. 
The important point, however, is that it is usually possible with 
the signal-to-noise ratio and resolution of our data to assess the degree 
of saturation of the Zn~II and Cr~II lines.

Column (4) lists the ratios $N$(Zn$^+$)/$N$(H$^0$) derived by dividing 
the entries in column (3) by those in column (3)
of Table 4; comparison with the solar abundance of 
Zn, log~(Zn/H)$_{\sun} = -7.35$ (Anders \& Grevesse 1989), then leads 
to {\it underabundances} of Zn by the factors given in column (5). The 
corresponding values for Cr (log~(Cr/H)$_{\sun} = -6.32$) are given in 
column (8) and column (9) lists the ratio $N$(Cr$^+$)/$N$(Zn$^+$) in 
cases where it could be determined.

In taking the ratios $N$(Zn$^+$)/$N$(H$^0$) and $N$(Cr$^+$)/$N$(H$^0$)
as measures of (Zn/H) and (Cr/H), we implicitly assume that there is
little contribution to the observed Zn~II and Cr~II absorption from
ionised gas (which would not produce \lya\ absorption).
This is likely to be the case given the large column densities of H~I,
and indeed there are no indications to the contrary in our data. In
particular, we found no significant differences in redshift between
the Zn~II and Cr~II lines, when detected, and O~I~$\lambda 1302.1685$
which arises only in H~I regions. Should this assumption be shown to
be incorrect, however, the values of [Zn/H] and [Cr/H] deduced here
would then be upper limits to the true abundances.

We now comment briefly on each DLA in Table 3.\\

\subsection{Q0000$-$263; $z_{\rm abs} = 3.3901$}

Our observations of this DLA, the highest redshift absorber in the
survey, have been described in Pettini et al. (1995a).
While Zn~II~$\lambda 2025.483$ remains undetected, despite the sensitive 
limit reached in a total exposure time of 58\,200~s,
we do record weak Cr~II absorptions at the $4\sigma$ 
($\lambda 2055.596$) and $3\sigma$ ($\lambda 2061.575$) significance levels.
Cr~II~$\lambda 2055.596$ is expected 
to be stronger than Zn~II~$\lambda 2025.483$
if the fraction of Cr locked up in dust grains is less than about 50\%.
With $N$(H$^0$) = ($2.5\pm0.5$)$ \times 10^{21}$~cm$^{-2}$ 
(Savaglio, D'Odorico, \& Moller 1994), this is one of the highest column 
density systems in our sample. We conclude that the abundance of Zn is 
less than 1/80 of the solar value; this estimate is $\sim 5$ times
more sensitive than the previous limit 
(Savaglio et al. 1994).  The abundance of Cr, [Cr/H] $\simeq -2.2 \pm 0.1$,
is similar to those of other elements measured by Molaro et al. (1996) and 
Lu et al. (1996), making  
this DLA one of the most metal-poor in our sample.\\

\subsection{Q0056$+$014; $z_{\rm abs} = 2.7771$}

This QSO is from the Large Bright Quasar Survey by Chaffee et al. (1991).
We deduce $N$(H$^0$) = ($1.3\pm0.2$)$ \times 10^{21}$~cm$^{-2}$  
from fitting the core of the damped \lya\ line, in reasonable
agreement with the value log~$N$(H~I) = 21.0 reported by Wolfe et al. 
(1995).

As can be seen from Figure 1, the Zn~II and Cr~II absorption lines are 
broad and shallow in this DLA, spanning $\approx 200$~km~s$^{-1}$. 
The stronger member of the Zn~II doublet, 
$\lambda 2025.483$, falls within the atmospheric A band.
Plotting the four 
absorption lines labelled in Figure 1 on the same velocity scale suggests 
that most of feature ``1'' is {\it not} due to Zn~II$~\lambda 2025.483$, but 
rather to poorly corrected telluric absorption. 
From the equivalent widths of Cr~II~$\lambda 2055.596$ and 
$\lambda 2065.501$ (features 2 and 4 in Figure 1), 
which are consistent with the optically thin ratio of 2:1, 
we deduce a weighted mean 
$N$(Cr$^+$) = ($2.8\pm0.4$)$ \times 10^{13}$~cm$^{-2}$.
This column density of Cr$^+$ produces an equivalent width 
$W_0 = $($82 \pm 12$)~m\AA\ for Cr~II~$\lambda 2061.575$; 
since we measure $W_0 = $($117 \pm 16$)~m\AA\
for feature 3, which is a blend of Cr~II$\lambda 2061.575$ and
Zn~II$\lambda 2062.005$, we conclude that 
$W_0 = $($35 \pm 20$)~m\AA\ for the latter.
This in turn corresponds to
$N$(Zn$^+$) = ($3.5\pm2$)$ \times 10^{12}$~cm$^{-2}$.
Thus both Zn and Cr appear to be $\approx 20$ times less abundant than 
in the Sun.

Our red spectrum also shows several Fe~II lines from an absorption system
at $z_{\rm abs} = 2.3044$, including: 
Fe~II~$\lambda 2344.214$ (visible in 
Figure 1 at $\lambda_{\rm obs} = 7748.46$\AA) with $W_0 = $($470 \pm 
12$)~m\AA;
Fe~II~$\lambda 2367.5905$ with $W_0 = $($64 \pm 6$)~m\AA;
Fe~II~$\lambda 2374.4612$ with $W_0 = $($220 \pm 14$)~m\AA; and
Fe~II~$\lambda 2382.765$ with $W_0 = $($640 \pm 12$)~m\AA.\\

\subsection{Q0201$+$365; $z_{\rm abs} = 2.462$}

Keck observations of this DLA have been published recently by Prochaska \& 
Wolfe (1996) who deduced relatively high abundances of Zn and Cr, 
respectively $\sim 1/2$ and $\sim 1/8$ of solar. Evidently, even at 
redshifts as high as 2.5 some galaxies had already undergone 
significant chemical evolution and enriched their interstellar media in 
heavy elements to levels comparable with that of the Milky Way today.\\

\subsection{Q0302$-$223; $z_{\rm abs} = 1.0093$}

Lanzetta, Wolfe, \& Turnshek (1995) 
proposed this as a candidate DLA system on the basis of 
low-resolution {\it IUE} data; a subsequent UV spectrum secured with the 
Faint Object Spectrograph on the  {\it Hubble Space Telescope} confirmed 
that  $N$(H$^0$) = ($2.15 \pm 0.35$)$ \times 10^{20}$~cm$^{-2}$
(Pettini \& Bowen 1997).
Recent WHT observations of Zn~II and Cr~II lines  
by Pettini \& Bowen (1997) have shown the abundances to be 
1/3 and 1/8 of solar respectively. 
After subtraction of the QSO radial profile from {\it HST} WFPC2 images of  
the field, Le Brun et al. (1997) identified two galaxies 
which may be producing the absorption; 
at $z = 1.009$ they would have luminosities 
$L \approx 0.2 L^{\ast}$ and $\approx L^{\ast}$ and
distances of 12 and $27~h_{50}^{-1}$~kpc respectively 
from the QSO sight-line.\\

\subsection{Q0454$+$039; $z_{\rm abs} = 0.8596$}

The abundances of Zn and Cr reported by Steidel et al. (1995a)
correspond to [Zn/H] = $-0.83 \pm 0.08$ and [Cr/H] = $-1.01 \pm 0.05$
if the experimentally measured $f$-values of the Zn~II and Cr~II multiplets 
(Bergeson \& Lawler 1993) are adopted for consistency with the rest of 
the present study. 
Deep images of the QSO field both from the ground (Steidel et al. 1995a)
and with {\it HST} (Le Brun et al. 1997)
suggest that the absorber is a compact galaxy with
$L \approx 0.25 L^{\ast}$ ($q_0 = 0.05$)
at a projected distance of $8~h_{50}^{-1}$~kpc from the QSO.\\

\subsection{Q0836$+$113; $z_{\rm abs} = 2.4651$}

This is the faintest QSO in our survey (Hunstead, Pettini, \& Fletcher 
1990) and the S/N of the red spectrum remains modest despite the 
considerable investment in exposure time (Table 1).
Combined with the relatively low H~I column density of
($3.8\pm0.4$)$ \times 10^{20}$~cm$^{-2}$,
the $3 \sigma$ upper limits to the Zn~II and Cr~II lines place limits on 
the abundances of Zn and Cr which are less stringent than in most other
DLAs considered: [Zn/H] $\leq -0.8$ and [Cr/H] $\leq -1.2$\,.

The blue spectrum shown in Figure 2 was recorded with the IPCS on the WHT  
in March 1994. Note that, of all the damped \lya\ lines reproduced in 
Figure 2, this is the only instance where there appears to be weak 
emission in the core of the absorption line. 
The line flux, ($2 \pm 0.7$)$ \times 10^{-17}$~erg~s$^{-1}$~cm$^{-2}$,
%corresponds to a 
%\lya\ luminosity 
%(0.9$\pm 0.3$)$ \times 10^{42}$~erg~s$^{-1}$ 
%($H_0 = 50$~km~s$^{-1}$~Mpc$^{-1}$; $q_0 = 0.5$) which
agrees within the errors with the value
of ($2.9 \pm 0.7$)$ \times 10^{-17}$~erg~s$^{-1}$~cm$^{-2}$ 
%(1.2$\pm 0.3$)$ \times 10^{42}$~erg~s$^{-1}$ 
reported by Hunstead et al. (1990) 
from independent data obtained in April 1987 with a different 
IPCS detector on the AAT. The two sets of observations
were obtained with the same slit width (1.2 arcsec) and at the same 
position angle on the sky (150 degrees).\\

\subsection{Q0841$+$129; $z_{\rm abs} = 2.3745, 2.4764$}

The spectrum of this bright ($V \simeq 17$), high redshift ($z \simeq 2.5$, 
estimated from the onset of the \lya\ forest) BL Lac object discovered 
by C. Hazard (private communication) shows {\it two} DLAs (see Figure 2),
making it a highly suitable target for follow-up high resolution 
observations. 

As can be seen from Figure 1, 
in the lower redshift system we detect features 2 and 3;  
the strength of the latter indicates a 
significant contribution from Zn~II~$\lambda 2062.005$ to the blend.
Following a procedure similar to that described for 
Q0056$+$014 at \S3.1 above, we deduce 
$N$(Cr$^+$) = ($9.5\pm2$)$ \times 10^{12}$~cm$^{-2}$
from the equivalent widths of Cr~II~$\lambda 2055.596$ and
$\lambda 2065.501$. This in turn leads us to estimate
that approximately half of the equivalent width of feature 3 is due to 
Zn~II~$\lambda 2062.005$ with $W_0 = $($24 \pm 9$)~m\AA.
Together with the $3 \sigma$ upper limit
$W_0$(2025)$ \leq 26$~m\AA\ for the stronger member of the doublet, this
then implies  
$N$(Zn$^+$) = ($1.8 \pm 0.5$)$ \times 10^{12}$~cm$^{-2}$.

Thus we find that Zn and Cr at $z_{\rm abs} = 2.3745$
are underabundant by factors of 23 and 45 respectively,
relative to solar values. 
Similar, or lower, abundances apply to the $z_{\rm abs} = 2.4764$ DLA, 
given the lack of detectable Zn~II and Cr~II lines 
(see Table 3).\\

\subsection{Q0913$+$072; $z_{\rm abs} = 2.6183$}

The signal-to-noise ratios of 
our spectra of this bright QSO are among the highest in the survey---see 
Table 1 and Figures 1 and 2.
The column density of neutral hydrogen is however comparatively low, 
$N$(H$^0$) = ($2.3 \pm 0.4$)$ \times 10^{20}$~cm$^{-2}$. The lack of 
Zn~II and Cr~II absorption even at S/N = 46 implies underabundances by 
factors of more than 14 and 32 respectively.\\

\subsection{Q0935$+$417; $z_{\rm abs} = 1.3726$}

Lanzetta et al. (1995) estimated 
$N$(H$^0$)$ \simeq 2 \times 10^{20}$~cm$^{-2}$
for this candidate DLA from low resolution {\it IUE} data; a subsequent
{\it HST} FOS spectrum confirmed that
$N$(H$^0$) = ($2.5 \pm 0.5$)$ \times 10^{20}$~cm$^{-2}$
(Lanzetta \& Meyer 1996, private communication).
With this value of the hydrogen column density,
the observations by Meyer, Lanzetta, \& Wolfe (1995)
imply [Zn/H] = $-0.80$ and [Cr/H] = $-0.90$\,.\\

\subsection{Q1104$-$180; $z_{\rm abs} = 1.6616$}

Smette et al. (1995) identified this DLA in the spectrum of the 
brighter ($B = 16.7$) component of this gravitationally lensed QSO 
pair.
From AAT observations obtained with an
instrumental setup similar to that used in our survey,
these authors estimated 
$N$(H$^0$)$ = 6 \times 10^{20}$~cm$^{-2}$. 
They also reported 
detections of Zn~II and Cr~II absorption lines with equivalent widths
$W_0$(2025.483)$ = {\rm (}75 \pm 20$)~m\AA\
and $W_0$(2055.596)$ = {\rm (}57 \pm 20$)~m\AA\ respectively.
If the lines are unsaturated
%, as suggested by the complex velocity 
% structure of other ions in this system, 
% then 
[Zn/H] = $-0.80$ and [Cr/H] = $-1.30$\,.\\

\subsection{Q1151$+$068; $z_{\rm abs} = 1.7736$}

Even though the damped \lya\ line falls in the crowded near-UV spectrum 
of this $z_{\rm em} = 2.762$ QSO (see Figure 2), our estimate
$N$(H$^0$) = ($2.0\pm 0.5$)$ \times 10^{21}$~cm$^{-2}$
is in very good agreement with log~$N$(H~I)$ = 21.3$ published by 
Turnshek et al. (1989).
The ratios of equivalent widths within the Zn~II and Cr~II multiplets 
strongly suggest that the lines are optically thin; 
Zn and Cr are both underabundant by a factor $\approx 40$.

Our red spectrum, which covers the region $5500-5900$~\AA, 
shows three C~IV~$\lambda\lambda1548,1550$ doublets at
$z_{\rm abs} = 2.5629$, 2.7069 and 2.7551 respectively.\\

\subsection{Q1209$+$093; $z_{\rm abs} = 2.5843$}

This is another high column density DLA; we measure 
$N$(H$^0$) = ($2.0\pm 0.5$)$ \times 10^{21}$~cm$^{-2}$
which compares well with log~$N$(H~I)$ = 21.4$
reported by Lu et al. (1993).
The Zn~II and Cr~II lines are the strongest
encountered in the entire survey of 34 DLAs (see Table 2).
Fitting the absorption profiles requires 
$b = 50$~km~s$^{-1}$ (as usual, $b = \sqrt{2}\sigma$ where
$\sigma$ is the one-dimensional velocity dispersion along the line of 
sight), indicating that several velocity components
most likely contribute to the absorption. 
Similarly, Lu et al. found 
$b = 122$~km~s$^{-1}$ from fitting a single curve-of-growth to the 
strongest UV absorption lines.
Some of the components may well be saturated; so we quote our best 
estimates of $N$(Zn$^+$) and $N$(Cr$^+$) as lower limits.
We conclude that Zn is {\it more} abundant than 1/9 solar and Cr more 
abundant than 1/27 solar.
Higher resolution observations are required to establish 
how close to these limits the true values are.\\

\subsection{Q1328$+$307; $z_{\rm abs} = 0.6922$}

We have included here the measurements of Zn and Cr abundances 
in the spectrum of 3C~286 reported by Meyer \& York (1992), after 
appropriate rescaling with the $f$-values of 
Bergeson \& Lawler (1993). 
Although the measurement was discussed in Pettini et al. 
(1994), this intermediate redshift DLA was not explicitly included 
in that sample which consisted exclusively of DLAs at 
$z_{\rm abs} > 1.7$\,.
CCD images of the QSO field obtained with ground-based telescopes 
(Steidel et al. 1994)
and with {\it HST} (Le Brun et al. 1997)
show a large ($\approx 10-30~h_{50}^{-1}$~kpc), 
low surface brightness galaxy which has been proposed as the absorber.\\

\subsection{Q1337$+$113; $z_{\rm abs} = 2.7957$}

Our measured 
$N$(H$^0$) of ($8 \pm 2$)$ \times 10^{20}$~cm$^{-2}$
agrees very well with 
log~$N$(H~I)$ = 20.9$ reported by Turnshek et al. (1989).
When we
observed this QSO, in March 1994, we found it to be considerably
fainter than the magnitude $V \simeq 18.2$ estimated by Hazard et al. 
(1986) from POSS plates.
Although the S/N achieved is the lowest in the survey
(see Table 1 and Figure 1), it is still sufficient
to establish that the abundances of Zn and Cr are less than 
1/10 and 1/23 of solar respectively.\\

\subsection{Q1946$+$769; $z_{\rm abs} = 2.8443$}

This $z_{\rm em} = 3.051$ QSO, intrinsically one of 
the most luminous known, is sufficiently bright to have been studied 
extensively at echelle resolutions and high S/N with 4-m telescopes
(Fan \& Tytler 1994; Lu et al. 1995; Tripp, Lu, \& Savage 1996).
However, the hydrogen column density
in the $z_{\rm abs} = 2.8443$ DLA is relatively low, 
$N$(H$^0$) = ($2 \pm 0.5$)$ \times 10^{20}$~cm$^{-2}$ (Lu et al.). 
Consequently, the upper limits [Zn/H] $\leq -0.82$ and
[Cr/H] $\leq -1.00$ 
deduced by these authors 
are rather uninformative given that the true metallicity
is  $\sim30$ times lower ([Fe/H]$ = -2.44 \pm 0.13$).\\

\subsection{Q2239$-$386; $z_{\rm abs} = 3.2810$}

This QSO is faint and the absorber is at high redshift; the combination of 
these two factors resulted in the longest integration time in 
the survey (see Table 1).
Adopting $N$(H$^0$)$ = 5.8 \times 10^{20}$~cm$^{-2}$
measured by Lu \& Wolfe (1994),
we deduce Zn and Cr underabundances by factors of more than 11 and 13 
respectively.

The Cr measurement is based on the weakest member of the triplet,
Cr~II$~\lambda 2065.501$;  
$~\lambda 2061.575$ is affected by a strong sky emission line
and $~\lambda 2055.596$, 
which at $z_{\rm abs} = 3.2810$ is redshifted to 
$\lambda_{\rm obs} = 8802.82$~\AA,
falls very close to Mn~II~$\lambda 2606.462$ at
$z_{\rm abs} = 2.3777$, the redshift of a second
DLA along this line of sight (Lu \& Wolfe 1994).
Based on the strengths of the other two members of the Mn~II triplet,
$\lambda 2576.877$ at $\lambda_{\rm obs} = 8703.55$~\AA\
and $\lambda 2594.499$ at $\lambda_{\rm obs} = 8763.97$\AA,
the feature labelled 2 in the last panel of Figure 1 is 
mostly Mn~II$~\lambda 2606.462$\,.
The two strong absorption lines also
visible in this figure 
are Fe~II~$~\lambda\lambda 2586.6500,2600.1729$
at $z_{\rm abs} = 2.3777$\,.\\

\section {DISCUSSION}

Our total sample, which consists of measurements (or upper limits) 
of [Zn/H] in 34 DLAs over the redshift range
$z_{\rm abs} = 0.6922 - 3.3901$,
is constructed by combining data for the 17 DLAs in Table 3
with those for the 15 DLAs in Table 3 of 
Pettini et al. (1994) and with the further addition of 
two DLAs in Q0528$-$250 (Meyer et al. 1989) which were included in 
the sample considered by Pettini et al. (1994) but not listed in their 
Table 3. All the points in Figure 3 are based on the 
$f$-values of the Zn~II doublet by Bergeson \& Lawler (1993) and the 
meteoritic solar abundance of Zn from the compilation by Anders \& Grevesse 
(1989)\footnote{This set of atomic parameters and solar abundance introduces
a correction of $-0.148$ to the values of [Zn/H] published in Pettini
et al. (1994)}.

We now consider what implications can be drawn from this extensive survey
on the chemical evolution of the neutral content of the universe and on 
the relationship of damped \lya\ systems to present-day spiral galaxies.\\

\subsection {Chemical Evolution of Damped \lya\ Systems}

Figure 3 shows the abundance of Zn as a function of redshift.
The enlarged sample confirms the two main conclusions reached by Pettini 
et al. (1994): 

(1) Damped \lya\ systems, {\it at all redshifts 
probed}, are generally metal-poor and presumably arise in galaxies at 
early stages of chemical evolution.

(2) There appears to be a large range in the values of metallicity 
reached by different galaxies at the same redshift, pointing to 
a protracted `epoch of galaxy formation' and to the fact that
chemical enrichment probably proceeded at different rates in different 
DLA galaxies. 

While we find gas with near-solar metallicities
at redshifts as high as $z \simeq 2.5$, there are also examples
of galaxies with abundances less than 1/10 solar at a time when the 
disk of the Milky Way differed little from its present-day composition. 
At redshifts $z \simeq  2 - 2.5$ the full range of metal
abundances spans about two orders of magnitude. 
Although for metallicities $Z_{\rm DLA} \simlt 1/50 Z_{\sun}$
the Zn~II lines become vanishingly small and only upper limits to 
the abundance of Zn can be deduced, 
we do know from echelle spectroscopy of more 
abundant astrophysical elements that values of  
$Z_{\rm DLA} \simlt 1/100 Z_{\sun}$  are not uncommon at 
$z_{\rm abs} = 2 - 3$ (see Figure 1 of Pettini et al. 1995a).

These two results are considered quantitatively in Table 5 where in 
the last column  we list,
for various subsets of our sample,  
the column density-weighted mean abundance of Zn 
\begin{equation}
	{\rm [} \langle{\rm Zn/H}_{\rm DLA}\rangle {\rm ]} = 
        {\rm log}\langle{\rm(Zn/H)}_{\rm DLA}\rangle-{\rm log~(Zn/H)}_{\odot}, 
	\label{}
\end{equation}
where
\begin{equation}
	\langle{\rm (Zn/H)}_{\rm DLA}\rangle = 
        \frac{\sum\limits_{i=1}^{n} N{\rm(Zn}^+{\rm)}_i}
        {\sum\limits_{i=1}^{n} N{\rm(H}^0{\rm)}_i} , 
	\label{}
\end{equation}

\noindent and $\sigma^{\prime}$, 
the standard deviation from the column density weighted mean, defined as

\begin{equation}
	 (n - 1)~ \sigma^{\prime 2} = 
        \sum\limits_{i=1}^{n}
        \lgroup {\rm[Zn/H]}_i - 
         {\rm [} \langle{\rm Zn/H}_{\rm DLA}\rangle {\rm ]}\rgroup ^2
	\label{}
\end{equation}

The summations in equations (2) and (3) are over 
the $n$ DLA systems considered in each subsample.

Under the working assumption that DLAs account for most of the material 
available for star formation at high redshift,
the quantity ${\rm [} \langle{\rm Zn/H}_{\rm DLA}\rangle {\rm ]}$
is a measure of the degree of metal enrichment reached by the universe at 
a given epoch. This is a general statement which 
follows from the column density distribution of \lya\ systems
(Lanzetta et al. 1995) and which holds
irrespectively of the precise 
nature of the damped absorbers (disks, spheroids, 
gas clouds yet to collapse into galaxies, etc.),
provided that there are 
no significant biases in the samples of DLAs from which our observations 
are drawn (Fall 1996).
        
The values of ${\rm [} \langle{\rm Zn/H}_{\rm DLA}\rangle {\rm ]}$
in Table 5 are strictly upper limits (with the exception of subsample 
number 1), since the averages include systems for which 
only upper limits to the abundance of Zn are available.
However, we expect the corrections to be small because 
the systems where the Zn~II doublet is below our detection limits are 
preferentially those with the lowest values of hydrogen column density 
$N$(H$^0$). Specifically, 
the fractions of $\sum\limits_{i=1}^{n} N{\rm(H}^0{\rm)}_i$ contributed by
DLAs with undetected Zn~II lines are 28\% for the full sample, 
and 16\%, 16\%, and 37\%
respectively for subsamples 2, 3 and 4. 
To show that including the upper limits as detections has only a modest
effect on the mean values of metallicity deduced, 
we have recalculated 
${\rm [} \langle{\rm Zn/H}_{\rm DLA}\rangle {\rm ]}$ for the full
sample twice, substituting $2\sigma$ and $1\sigma$ limits respectively 
in place of the
$3\sigma$ limits used in Table 5 (it could indeed be argued that 
$3\sigma$ limits for the entire ensemble on Zn non-detections is an
overly conservative approach). 
In this case, ${\rm [} \langle{\rm Zn/H}_{\rm DLA}\rangle {\rm ]}$
decreases from $-1.13 \pm 0.38$ (the value listed in Table 5) to
$-1.16 \pm 0.40$ and $-1.20 \pm 0.48$ respectively.
On the other hand,  
{\it all} three measurements in subsample 5 
($z_{\rm abs} = 3.0 - 3.5$)
are upper limits and accordingly we quote the value of  
${\rm [} \langle{\rm Zn/H}_{\rm DLA}\rangle {\rm ]}$
in this redshift interval
as an upper limit.

For the full sample of 34 DLAs in the range $z_{\rm abs} = 0.7 - 3.4$
we find ${\rm [} \langle{\rm Zn/H}_{\rm DLA}\rangle {\rm ]} = -1.13 \pm 0.38$.
This is the same value as obtained by Pettini et al. (1994)
when account is taken of the different $f$-values 
and solar abundance scale used in our earlier study.
For comparison, [Zn/H]$_{\rm gas} = -0.19$ along 
unreddened sight-lines in the solar vicinity 
(Roth \& Blades 1995; Sembach et al. 1995---both analyses used the
same $f$-values and solar scale as here).
If the interstellar medium (gas+dust) near the Sun has the same 
composition as the Sun, this would imply that approximately 35\% of Zn 
is in solid form. 
On the other hand, Pettini et al. (1997) found that for 
the present sample of DLAs the typical 
dust-to-metals ratio is 
approximately half that of the Galactic ISM. 
If we assume, therefore, that on average 83\% of Zn in DLAs is in 
the gas phase, we obtain
${\rm [} \langle{\rm Zn/H}_{\rm DLA}\rangle {\rm ]} 
= -1.13 \pm 0.38 + {\rm log (}1/0.83{\rm )} = 
-1.004 \pm 0.38$, and conclude that the
column density weighted abundance of Zn in DLAs is 1/11 of that of the Milky 
Way ISM today.

One of the motivations of the present work was to 
determine the redshift evolution of 
the metallicity of DLAs and thereby 
trace the increase of heavy elements in 
the universe from the epoch of galaxy formation to the present time.
From Figure 4, where our measures of 
${\rm [} \langle{\rm Zn/H}_{\rm DLA}\rangle {\rm ]}$ from 
Table 5 are plotted versus redshift, 
it can be seen that any such evolution is 
only mild in the present sample.
Between $z = 3$ and 1.5, to which 80\% of the sample refers,
there appears to be little change from the typical 
${\rm [} \langle{\rm Zn/H}_{\rm DLA}\rangle {\rm ]} = -1.13$\,.
This is less surprising, however, when one considers that this redshift 
interval spans a period of only $\approx 3$~Gyr
from 14.3 to 11.4 Gyr ago
($H_0 = 50$~\kms\  Mpc$^{-1}$;  $q_0  =  0.01$)
and that at these epochs evidently there was a large spread in the 
chemical enrichment of different DLA galaxies.

On the other hand, the upper limit 
${\rm [} \langle{\rm Zn/H}_{\rm DLA}\rangle {\rm ]} \leq -1.39$ at 
$z > 3$ is lower than the means in the other redshift bins,
providing tentative evidence for 
a rapid build-up of elements with time at this epoch.
This suggestion is strengthened by the data of Lu et al. (1996)
who found that [Fe/H] \simlt $-2$ in three additional DLAs at 
$z > 3$\,.  (The correction to [Fe/H] for the fraction of Fe
in solid form is likely to be small---probably less than a factor of 
two---at such low metallicities; 
Pettini et al. 1997).
% Qualitatively, this trend is already visible in 
% Figure 3 where, despite having three times the number of DLAs 
% with $z_{\rm abs} > 2.5$ compared with our earlier sample, 
% we have found no 
% new cases with abundances higher than $\sim 1/10$ solar.
The lowest metallicities measured in DLAs,
$Z_{\rm DLA} \simeq -2.5$,
are comparable to those thought to apply to  
the ionised intergalactic medium producing the \lya\ forest 
at redshifts $z = 2 - 3.5$
(Hellsten et al. 1997), although the large ionization corrections 
involved make estimates of $Z_{\rm IGM}$ considerably more 
uncertain than $Z_{\rm DLA}$.
It is tempting, therefore, to interpret the rapid increase in 
metal abundances at $z < 3$ as an indication of 
the onset of star-formation in galaxies
and to speculate that $Z \simeq -2.5$ may be an approximate 
`base' level of metallicity
on which galactic chemical evolution subsequently builds.

The recently realised ability to image high redshift galaxies directly 
in their ultraviolet stellar continua has led to the first attempts to 
sketch the global history of star formation over $\sim 80$\% of the age 
of the universe 
(Madau et al. 1996 and references therein). 
Determinations of the volume-averaged star formation rate (SFR) from
the so-called $B$ and $U$ drop-outs 
(galaxies with the Lyman limit in the $B$ and $U$ bands respectively) 
in ground-based surveys (Steidel, Pettini, \& Hamilton 1995b) and 
in the {\it Hubble Deep Field} (Madau 1996) do indeed suggest an increase
in the SFR from $z \simeq 4$ to $z \simeq 2.75$\,. 
As discussed by Madau et al., 
it is possible to convert the integrated UV luminosity density 
into a metal ejection rate $\dot{\rho_Z}$
per comoving volume at redshift $z$.
Since the massive stars which are the main contributors to the 
far-UV continuum are also the major producers of heavy elements 
(at least those released into the ISM by Type II supernovae),
the conversion does not depend sensitively on the shape of the IMF
in these primordial galaxies. Rather, the principal sources of 
uncertainty arise from the cosmology assumed and from the amount of dust 
extinction suffered by the UV continuum.

Bearing in mind these uncertainties, it is of great interest to compare
the values of $Z_{\rm DLA}$ deduced here with the metallicities which may 
be expected on the basis of Madau's metal ejection rate. 
Integrating $\dot{\rho_Z}$ in Figure 3 of Madau (1996) from $z = 5.5$ to 
the present time yields a total density of metals 
$\rho_Z{\rm (}z = 0{\rm )} \approx 6.2 \times 10^6 M_{\sun}$~Mpc$^{-3}$.
This corresponds to an approximately solar metallicity 
if the present day density of baryons in galaxies is 
$\rho_{\ast}{\rm (}z = 0{\rm )} \approx {\rm (}2.7 \pm 0.4{\rm )} \times 10^8 
M_{\sun}$~Mpc$^{-3}$, or $\Omega_{\ast} \approx 4 \times 10^{-3}$  
($H_0 = 50$~km~s$^{-1}$~Mpc$^{-1}$; Madau et al. 1996).
In a closed box model, assuming that  
$\Omega_{\ast}(z = 0) \approx \Omega_{\rm DLA}(z = 4)$
(Storrie-Lombardi, McMahon, \& Irwin 1996a)
we can take 
\begin{equation}
	\frac{Z(z)}{Z(0)} \simeq \frac{Z(z)}{Z_{\sun}} =
	\frac{\displaystyle \int_{5.5}^{z} \dot{\rho_Z} dz}
             {\displaystyle \int_{5.5}^{0} \dot{\rho_Z} dz}
	\label{}
\end{equation}
provided the gas consumption into stars from $z = 5.5$ to $z$ is low and
$\Omega_{\rm gas}{\rm (}z {\rm)} \gg \Omega_{\ast}{\rm (}z {\rm)}$. 
The redshift evolution of $\Omega_{\rm DLA}$ 
(Storrie-Lombardi et al. 1996a) suggests that this may well 
be the case up to $z \simeq 2$ 
(as we proposed in Pettini et al. 1994).

The broken line in Figure 4 shows 
the increase of ${Z(z)}/{Z_{\sun}}$ with decreasing redshift
calculated from equation (4) and Madau's (1996)
estimates of $\dot{\rho_Z}$.
Evidently, there is rough agreement between the predicted and 
observed values of $Z_{\rm DLA}$.
Given the current 
uncertainties, we consider it premature to read too much 
into this comparison.
% and the agreement may be fortuitous to some extent.
For example, Madau's $\dot{\rho_Z}$ refers primarily to oxygen and the 
$\alpha$-elements which presumably are more abundant than zinc and iron
by a factor of $2-3$ at these low metallicities (Edvardsson et al. 1993; 
Carney 1996). On the other hand, the broken line in Figure 4 may well 
underestimate the metal production rate by similar factors if
star-forming galaxies at high redshift are reddened by small amounts of 
dust, corresponding to ${\rm E(B-V)} \approx 0.1$, as suggested by the 
observed slopes of the UV continua (Steidel et al. 1996). 

Nevertheless, taken at face value, Figure 4 does seem to indicate that in the 
DLAs we see roughly the same level of metal enrichment as expected 
from direct observations of star-forming galaxies at these redshifts.
More complex galactic chemical evolution models which use
as a starting point the 
gas consumption indicated by the redshift evolution of 
$\Omega_{\rm DLA}$ (Pei \& Fall 1995; Fall 1996)
also reproduce the degree of metal enrichment of DLAs
and the comoving rate of star formation at 
$z \simgt 2$\,.  Thus it appears that, to a first approximation at least,
these three independent avenues to exploring the epoch of galaxy 
formation---the consumption of neutral gas, the
metal abundance of the absorbers, and the UV luminosity 
of high-redshift galaxies---give a broadly consistent picture 
of the early evolution of galaxies.\\

\subsubsection{Abundances at $z < 1.5$}

The situation is less clear at lower redshifts.
Only four measurements make up subsample 1 in Table 5, 
even though this bin 
spans a larger interval of time than 
all the other subsets put together---$\approx 5$~Gyr
from 11.4 to 6.3 Gyr ago
(again for $H_0 = 50$~\kms\  Mpc$^{-1}$;  $q_0  =  0.01$).
Evidently 
${\rm [} \langle{\rm Zn/H}_{\rm DLA}\rangle {\rm ]} = -0.98 \pm 0.33$
is below an extrapolation of Madau's curve in Figure 4.
However, it is difficult to assess how firm this 
conclusion is, 
given that 65\% of $\sum\limits_{i=1}^{n} N{\rm(H}^0{\rm)}_i$
for subsample 1 is due to the $z_{\rm abs} = 0.6922$ absorber in 3C~286
which appears to be a large, low surface brightness galaxy 
(see \S3.13 above). 
Possibly such galaxies, 
whose low metallicities at the present time
are thought to be the result of low 
star formation efficiencies 
(McGaugh 1994; Padoan, Jimenez, \& Antonuccio-Delogu 1997),
come to dominate the cross-section for DLA absorption at 
$z \simlt 1$, if by this epoch most high surface brightness galaxies have  
already processed a significant fraction of their gas into stars.
Furthermore, the build-up of dust which goes 
hand-in-hand with the production of 
metals is likely to introduce an increasing bias (with decreasing redshift) 
against galaxies in advanced stages of chemical evolution, since
existing samples of damped \lya\ systems are mostly drawn from magnitude
limited optical QSO surveys (Fall \& Pei 1993; Pei \& Fall 1995).  

At $z_{\rm abs} < 1.5$ imaging of DLA absorbers is within current 
observational capabilities. 
Although positive identifications based on spectroscopic 
redshifts have not yet been achieved, 
the candidates which have been proposed 
suggest a very diverse population of galaxies.
While in some cases the
absorbers could be on 
evolutionary paths similar to that of the Milky Way, the 
$z_{\rm abs} = 1.0093$ DLA in Q0302$-$223 being a good example
(Pettini \& Bowen 1997),
there are also several instances where galaxies of low luminosity 
($L_B \simlt 0.1 L^{\ast}$) or of low surface brightness are indicated
(Steidel et al. 1994, 1995, 1997; Le Brun et al. 1997).

Thus both effects considered above---a shift of the DLA population away 
from `normal' $L^{\ast}$ galaxies and an increasing dust bias---may  
contribute to the finding that $Z_{\rm DLA}$ does not increase
significantly at 
$z_{\rm abs} < 1.5$ in Figures 3 and 4, 
contrary to simple expectations in a 
closed-box model of chemical evolution. 
However, it will not really be possible to proceed further, 
and quantify the relative importance of these two effects,
without a larger sample of DLAs at intermediate 
redshifts. Identifying such a sample remains an urgent priority.\\

\subsection {Comparison with Stellar Populations of the Milky Way}

Damped \lya\ systems are commonly thought of as the high redshift 
counterparts of present-day galactic disks, although we and others 
(Pettini et al. 1990; York 1988) have often made the point 
that high column densities of neutral gas are not the prerogative of disk 
galaxies alone. The sample of [Zn/H] measurements now available is 
sufficiently large to allow a comparison to be made of the  
distribution of metallicities in DLAs 
with those of different stellar populations in 
the Milky Way.  
In the solar cylinder, stars in the halo, thick 
disk, and thin disk have distinct dynamical and abundance properties, 
although the distributions overlap in either parameter taken separately. 
It is the {\it combination} of chemical abundance and kinematic data 
that studies of Galactic evolution have focussed on; here we attempt to 
use this information to throw light on the nature of DLA galaxies.

Our measurements of [Zn/H] from Figure 3 have been plotted in the top 
panel of Figure 5 after converting redshift to look-back time in a cosmology 
compatible with stellar ages. The lower panel in Figure 5 shows the 
age-metallicity relationship for disk stars determined in the landmark 
study by Edvardsson et al. (1993). This sample includes stars with the 
kinematics of both thin and thick disk, defined in terms of the 
mean velocity perpendicular to the Galactic plane:
$\langle |W| \rangle = 19$~\kms\ and $\approx 42$~\kms\ for thin and 
thick disk stars respectively (Freeman 1991).
In constructing their sample, Edvardsson et al. aimed to include 
approximately equal numbers of stars in each metallicity bin above 
$Z = 0.1 Z_{\sun}$; consequently, metal-poor stars are relatively 
over-represented in Figure 5.

Stellar ages are notoriously uncertain, as is the mapping of redshift to 
look-back time. However, even allowing for an arbitrary sliding 
of the points in Figure 5 along the $x$-axes, the metallicity measurements
in DLAs evidently do not match the chemical evolution of
the Milky Way disk.
The {\it typical} value $Z_{\rm DLA} = -1.13$ is lower than that of  
even the most metal-deficient stars in the Edvardsson et al. survey, 
and at all ages the spread of chemical abundances in the disk  
is smaller than that of the DLA sample.

This point is reinforced by Figure 6, where we compare the metallicity 
distribution of DLAs with those of stars in the thin disk, thick disk and 
halo populations; in Figure 7 we show the comparison with the metallicity
histogram for globular clusters. 
Values for disk stars are from the work by Wyse \& Gilmore (1995).
These authors combined spectroscopic determinations of [Fe/H] for a 
sample of F and G stars located $1-5$~kpc from the plane
with data for samples near the Sun, paying  
particular attention to including
only stars with potential main-sequence 
lifetimes greater than 12 Gyr. 
That is, their combined sample should be complete, 
in the sense of not missing
disk stars which have by now evolved away from the main sequence, and 
the resulting metallicity distributions presumably provide an 
integrated record of the chemical 
evolution of the disk. 
The thin disk distribution in Figure 6
includes the low metallicity tail discussed by Wyse \& Gilmore (1995);  
similarly, the thick disk histogram is consistent with the metal-weak 
tail shown in Figure 22 of Beers \& Sommer-Larsen (1995).
The halo sample is from the survey of high proper-motion 
stars in the solar neighbourhood by Laird et al. (1988), while
the histogram in Figure 7 is based on the distribution of [Fe/H] in 40 
globular clusters plotted in Figure 16 of Carney et al. (1996). 

The comparison between the metallicity distribution of DLAs and those of 
stellar populations in the Galaxy is complicated by the fact that about 
half of the values which make up the bins with [Zn/H]$_{\rm DLA} \leq -0.8$ 
in Figures 6 and 7 correspond to upper limits of [Zn/H] in our survey.
Were we to exclude the upper limits from the sample, 
the resulting distribution would be skewed 
to higher metallicities. 
This is also the case if they are included in 
the sample as detections, as we have done; 
therefore {\it the true distribution
of $Z_{\rm DLA}$ is both broader and shifted towards lower metallicities 
(by undetermined amounts) than the histogram reproduced in Figures 6 and 7}.

Bearing this in mind, the middle and bottom panels of Figure 6 show that
the metallicity distribution of DLA galaxies is different 
from those of long-lived stars in the Galactic disk.
Although there is some overlap with the thick disk histogram, the bulk of 
stars in the disk of the Milky Way apparently formed from gas which was 
significantly more metal-rich than that giving rise to damped \lya\ 
systems. 
The narrow distributions for disk stars in Figure 6
reflect the finding by Edvardsson et al. (1993) that the average metallicity
has increased very little over the lifetime of the disk;
the scatter at any age in the bottom panel of Figure 5 is nearly as
large as the difference in mean metallicity over the entire time span
considered.
This is also the case for the old open clusters of the Milky Way disk
(Friel 1995).

The width of the $Z_{\rm DLA}$ distribution is comparable to 
those of halo stars and globular clusters, but it peaks 
at a higher metallicity. 
This is probably a real effect, 
rather than being due to the inclusion of 
upper limits in our sample (as discussed above), 
since the column density weighted mean 
metallicity is ${\rm [} \langle{\rm Zn/H}_{\rm DLA}\rangle {\rm ]} = -1.13$\,.
We consider it unlikely that the offset between the observed and 
true peaks of the $Z_{\rm DLA}$ distribution is as 
large as required to bring the histograms in the top panel of Figure 6 and in 
Figure 7 into agreement.
Rather we favour the interpretation that, as a whole, 
the population of 
DLA galaxies is genuinely more metal enriched than the stellar components 
of the Galactic halo. 

The comparisons discussed above lead to two possible conclusions 
concerning the nature of damped \lya\ galaxies.
The most straightforward interpretation is that
a wide range of galaxy morphological types, at different stages of 
chemical evolution, make up the the DLA population.
Available imaging data at $z \simlt 1$ are certainly consistent 
with this view. 
A more intriguing possibility is that DLA systems at high redshift 
arise primarily in the spheroidal component of the present-day galaxy 
population, by analogy with the interpretation 
of the $U$ drop-out galaxies put forward by Steidel et al. (1995, 1996).
In the Milky Way, the halo and inner bulge may well be related, 
with the halo having lost $\approx 90$\% of its mass 
to the bulge (e.g. Wyse, Gilmore, \& Franx 1997); 
in this picture the halo-bulge system is an evolutionary 
sequence parallel to that of the thick disk-thin disk.
One could speculate, then, that the distribution of  
$Z_{\rm DLA}$, with its peak at a higher metallicity than halo stars
and globular clusters,
reflects different stages in the transition from metal-poor 
halo to a predominantly metal-rich bulge (Ibata \& Gilmore 1995).\\

\subsubsection{Divergent Clues from the Absorption Line Profiles?}

The message conveyed by Figure 6 contrasts with the interpretation 
by Wolfe and collaborators of the 
complex absorption line profiles, often extending over more than 100~\kms, 
revealed by high resolution spectroscopy
of DLAs  (Wolfe 1995; Prochaska \& Wolfe 1997b).
These authors have argued that in many cases the different  
components which make up the absorption lines
are not distributed at random in velocity;  
rather, there appears to be a more regular trend of decreasing
optical depth with increasing  
velocity difference from the wavelength where the absorption is strongest.
This `edge-leading asymmetry' is the pattern which would be produced by a 
rotating thick disk, intersected at some distance from the centre, 
if the average density of gas 
falls off with distance from the centre and from midplane.
Prochaska \& Wolfe show that the frequency
with which such absorption profiles are encountered 
is consistent with expectations for randomly oriented disks;
this leads them to conclude that most, if not all, DLAs arise in large
($R > 10$~kpc) disks with high rotation velocities 
($v_{\rm rot} \simgt 200$~\kms). 
Such structures, if common at $z \simgt 2$,
are very difficult to explain in currently favoured
models of galaxy formation (e.g. Baugh et al. 1997).
 
The Milky Way is the only galaxy for which we have a record of both 
chemical abundances and kinematics over its past history.
Based on this body of data, the metallicities we measure 
in the damped \lya\ systems 
appear incompatible with the rotating 
disk interpretation put forward by Prochaska \& Wolfe.
This can be appreciated by considering 
compilations of metallicities and velocities now available for 
large samples of stars, such as that published recently by 
Carney et al. (1996). 
From their Figures 1 and 3 it can be seen that, of the  
stars with metal abundances similar to those of DLAs, 
approximately half have {\it retrograde} motions;
at a metallicity $Z = -1.1$ 
the mean velocity relative to the disk rotation 
is $\langle {\rm V} \rangle \simeq -150$~\kms\,.  
This point is best illustrated by Figure 5 of Carney et al. which shows the 
metallicity histograms in various intervals of ${\rm V}$; 
our distribution of $Z_{\rm DLA}$ corresponds to values of 
${\rm V}$ in the range $\approx -100$ to $\approx -200$~\kms. 
Evidently, when our Galaxy had an average metallicity of 
$\simlt 1/10$ of solar, it did not exhibit the kinematics of a disk 
rotating at $\sim 200$~\kms.

Reconciling these contrasting clues to the nature of damped \lya\ galaxies
is an important task for the future. Here we put forward three 
possible ways out of the current impasse:

1. Our Galaxy is atypical, and the physical processes which gave rise to 
its stellar populations were not shared by most other galaxies at high 
redshifts. Although this possibility cannot be discounted, 
it is not a very constructive hypothesis to take refuge in,
as it will be difficult to test it 
observationally---at least in the near future.

2. The absorption profiles are being overinterpreted. 
A possible concern here
is that material whose motion is due 
not to rotation but to energetic events, such as supernova 
shocks, may contribute to the 
ultraviolet absorption lines, since these transitions
are sensitive to even 
relatively small column densities of gas.
% It is the case that not all absorption lines in a given DLA show the 
% kinematics of a rotating disk. 
% Rather, the pattern is usually best 
% discerned in a subset of lines with the appropriate range of optical 
% depths, but the choice
% may introduce an element of subjectivity
% in the analysis.
The `edge-leading asymmetry' interpretation was first proposed by 
Lanzetta \& Bowen (1992) in their analysis of  
13 Mg~II absorption components spread over 250~\kms\ in the 
$z_{\rm abs} = 0.39498$ DLA in Q1229$-$021. 
However, it is far from clear 
that this is really a massive disk; from their analysis of 
{\it HST} images of the field, Le Brun et al. (1997)   
propose that the absorber is instead a faint
($L_B < 0.1 L^{\ast}$) low surface brightness galaxy. 
Furthermore, strong Mg~II absorption spanning $\approx 300$~\kms\ can
also be produced by galaxies which are nearly face-on, such as M61
(Bowen, Blades, \& Pettini 1996).
All these factors cast some doubts on a detailed correspondence
between the profiles of ultraviolet absorption lines and the
large-scale kinematics of the intervening galaxies.

3. A third option, and one which we have already proposed, is that 
DLA galaxies comprise a mix of different morphological types.
Thus, it is conceivable that some do exhibit the kinematics of rapidly 
rotating disks, while others may be spheroids or 
irregular star-forming galaxies with less ordered velocity fields.
This is a hypothesis which {\it can} be tested. 
As more cases become available
where both kinematics and chemical abundances 
are measured in the same DLA,
it will be of great interest to examine
whether there is any correlation between these two parameters, as found
in the stellar populations of the Milky Way.\\

\section{SUMMARY AND SUGGESTIONS FOR FUTURE WORK}

We have assembled the largest sample of damped \lya\ systems for which 
metallicities have been measured free from the complications introduced 
by dust depletions. 
The expanded data set reinforces the two main conclusions reached 
in our earlier study (Pettini et al. 1994): (1) DLAs are generally 
metal-poor, {\it at all redshifts sampled}; and (2) there is a large 
spread in abundances at all epochs. We interpret these findings as 
evidence for a protracted epoch of galaxy formation, 
and propose that galaxies of different morphological types and at different
stages of chemical evolution make up the DLA population.

The metallicity distribution of DLAs is broader and peaks 
at lower metallicities than those of either the thin or thick disk of our 
Galaxy. Thus, the chemical abundance data presented here
do not support the interpretation of the absorption line profiles in 
terms of thick disks with rotation velocities 
$v_{\rm rot} \simgt 200$~\kms\ most 
recently discussed by Prochaska \& Wolfe (1997b).
This apparent discrepancy may be resolved by further work
on both the kinematics and the abundances. 
With the near-infrared spectrographs now being built for 8-10~m 
telescopes it will be possible to detect the 
familiar optical emission lines
from star-forming regions in the absorbing galaxies.
The widths of these features are likely to be more representative 
of the global kinematics than the 
ultraviolet absorption lines which can be so easily affected by 
local phenomena such as interstellar shocks.
On the abundance front, the ratios of chemical elements
manufactured in different nucleosynthetic processes
have been used to good effect in 
unravelling the history of star formation in our 
Galaxy; the same techniques are now beginning to be applied to 
high redshift DLAs (Pettini et al. 1995b; Lu et al. 1996).

The column density weighted mean metallicity of DLAs at $z \simgt 2$
is in agreement with  
expectations based on the metal ejection rate
deduced by Madau (1996) 
from the integrated ultraviolet 
luminosity of star forming galaxies at these redshifts.
Our data, when combined with the [Fe/H] measurements 
by Lu et al. (1996), 
appear to reflect the rapid increase in the comoving
star-formation rate between $z \approx 4$ and $\approx 2$
indicated by the relative numbers of $B$ and $U$ drop-outs in the 
{\it Hubble Deep Field}.
While these comparisons are of necessity still very approximate,
the implication seems to be that observations of DLAs 
provide a reasonably accurate census of  
metal enrichment at these epochs. 
It is encouraging that three independent methods which have been 
applied to the quest for
the epoch of galaxy formation---the global star formation rate 
deduced from the ultraviolet luminosity of high-redshift galaxies,
the rate of consumption of neutral gas
implied by the redshift evolution of $\Omega_{\rm DLA}$,  
and the metallicity of DLAs---apparently give a broadly consistent picture 
of the universe at $z \simgt 2$\,. 
 
This is not the case at $z \simlt 1.5$, where $Z_{\rm DLA}$ 
apparently does not 
rise as expected from 
simple models of cosmic chemical evolution.
There are a number of plausible explanations for this, including the
effects of dust, as discussed extensively by Fall and 
collaborators, and an increasing contribution of low surface brightness 
galaxies to the cross-section for DLA absorption. 
The major obstacle to progress in this area is still the paucity of 
DLAs with measured element abundances at intermediate redshifts.
And yet it is essential to follow the evolution of the DLA population to 
the present time in order to be confident of our interpretation of the 
high redshift data.
New DLAs at $z \simlt 1$ are still being identified and the sample is 
slowly growing. With STIS on the {\it HST} 
measurements of [Zn/H] can be extended to redshifts lower than the limit
$z_{\rm abs} \simeq 0.65$ of ground-based observations. 
In the next few years the 2dF and Sloan sky surveys
(Taylor 1995; Gunn \& Weinberg 1995) 
are expected to increase the number of known DLAs
by one order of magnitude.
With 8-10~m telescopes it will then be possible 
to repeat surveys such as this one towards substantially fainter, and 
potentially more reddened, QSOs.
Such programmes should lead to a better assessment 
of the significance of dust bias 
in current DLA samples.
Finally, with large telescopes we will soon be able to 
measure element abundances from the optical emission lines of galaxies at 
redshifts $z \approx  0.1-0.5$\,. Such data will complement in a very 
important way the information provided by galaxies selected from their 
absorption cross-section.\\

\acknowledgements
We are grateful to the  UK  and  Australian  Time  Assignment
committees  for generous  allocations  of  telescope  time  on
the  WHT and the AAT, and to the technical  staff  at  both
Observatories  for  excellent   support   with   the
observations.  The  WHT  Service  Observations  scheme  helped
bring this demanding observing programme to completion. We should
like to express our sincere thanks to: C. Hazard and P. Hewett 
for supplying us with positions and finding charts of QSOs with
candidate damped \lya\ systems; J. Lewis for assistance with the
reduction of some of the spectra; J. Laird and R. Wyse for
providing the stellar data used in Figures 6 and 7; C. Jenkins
and K. Lipman for help with some statistical aspects of the
analysis; M. Fall, G. Gilmore, P. Madau, and S. Ryan for
illuminating discussions on several issues relating to stellar
populations and galactic chemical evolution; and C. Steidel, J.
Prochaska, and D. York for useful comments on an earlier version
of the paper. R.W.H.  acknowledges financial assistance from the
Australian Research Council.

\newpage                                            
%\hspace*{-3.5cm}
\vspace*{-4cm}
\centerline{\epsfig{file=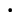,angle=180}}

\newpage
\vspace*{-4cm}
\centerline{\epsfig{file=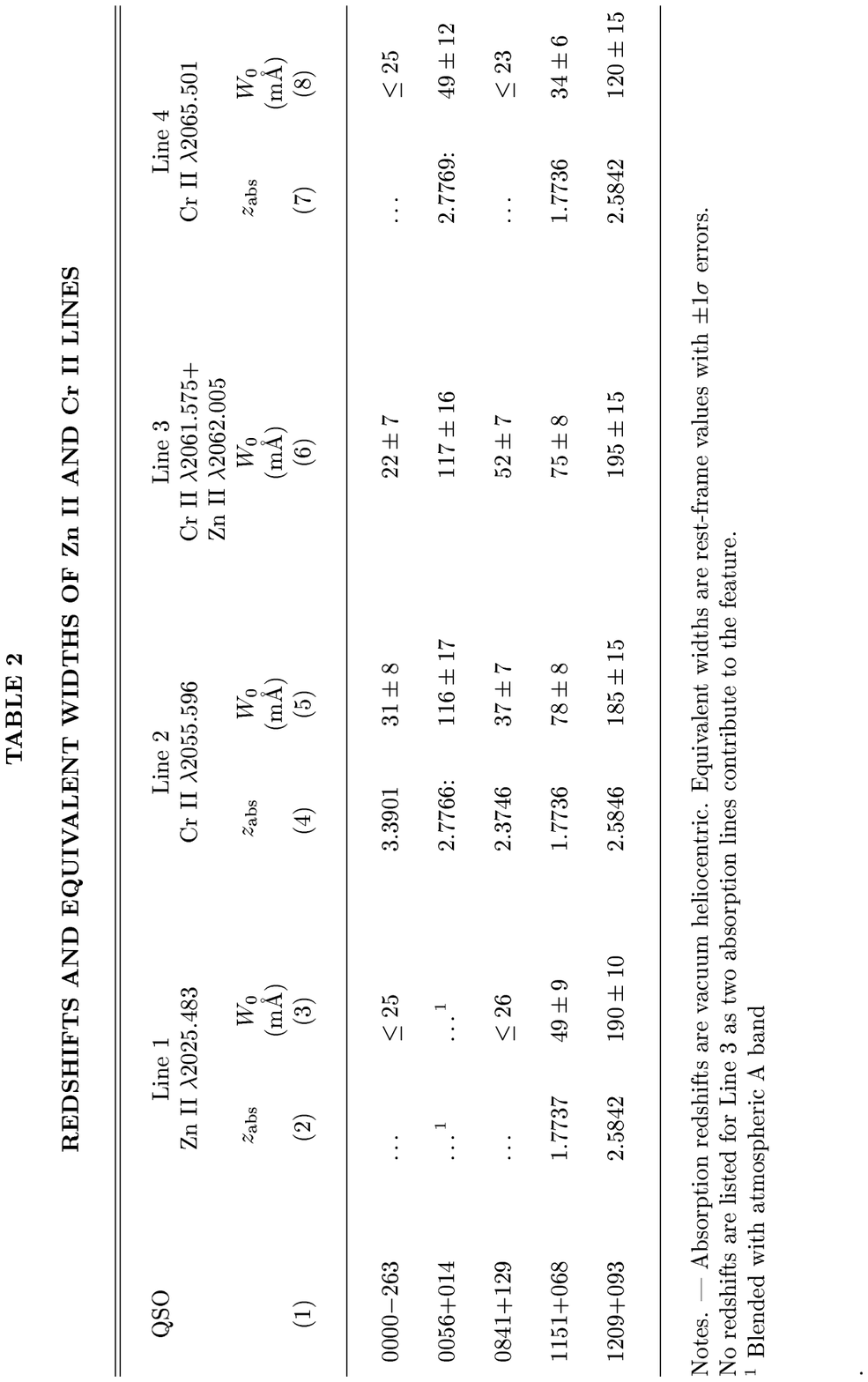,angle=180}}

\newpage
\vspace*{-1cm}
\centerline{\epsfig{file=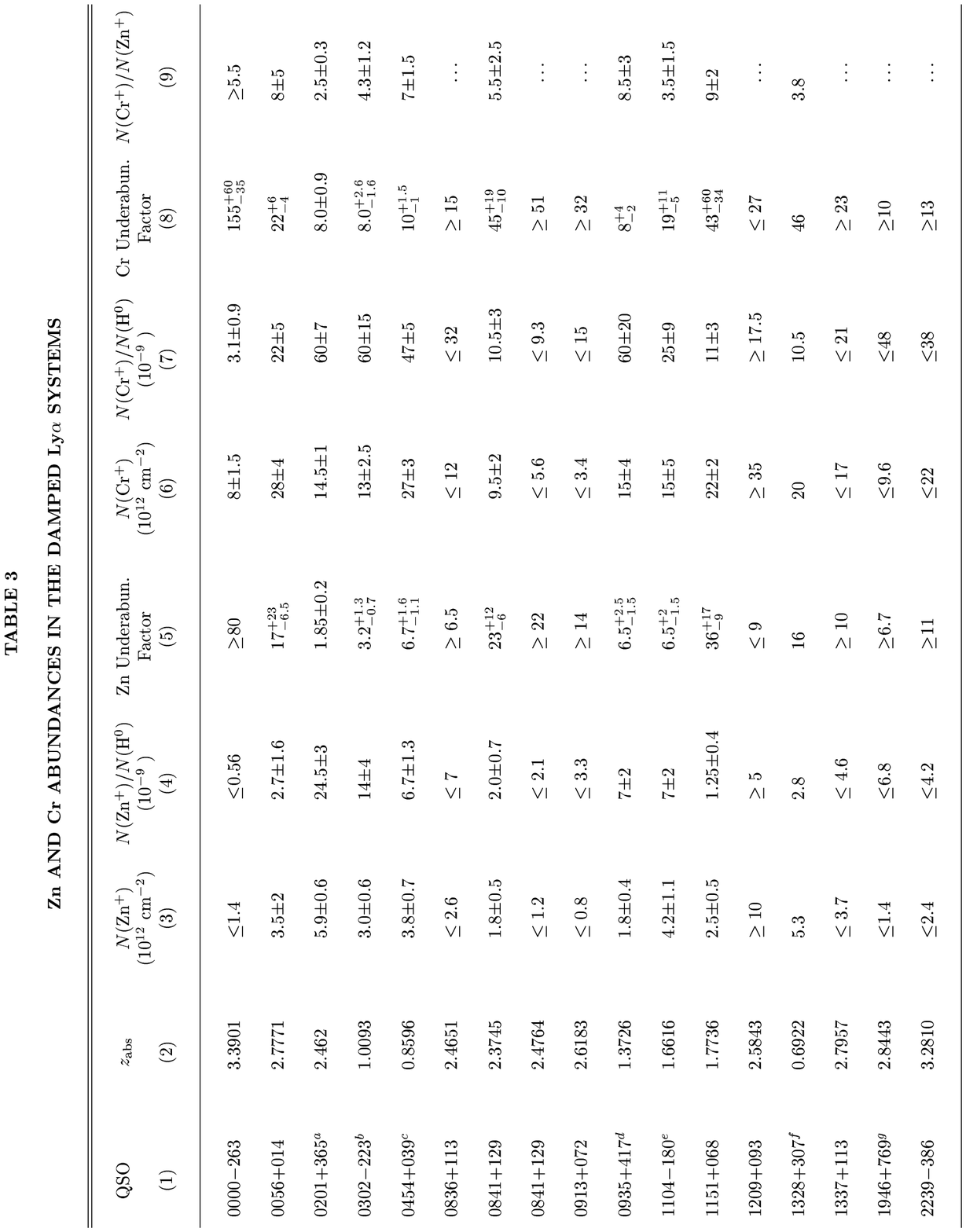,angle=180,width=18cm}}

\newpage
\hspace*{7cm}
\centerline{\epsfig{file=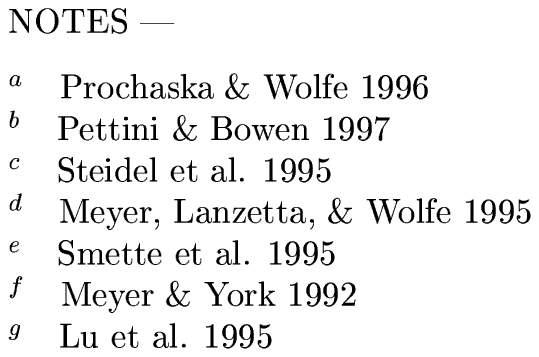,angle=90,width=35cm}}

\newpage
\vspace*{-4cm}
\centerline{\epsfig{file=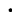,angle=0,width=19cm}}

\newpage
%\documentstyle[aj_pt4]{article}
%\documentstyle[apjpt4]{article}
%\begin{document}
%\addtocounter{page}{+4}
\addtocounter{table}{+4}
\begin{deluxetable}{lcccccc}
\tablewidth{0pc}
%\footnotesize
\scriptsize
\tablecaption{COLUMN DENSITY WEIGHTED METALLICITIES}
\tablehead{
\colhead{} & \colhead{Redshift Range} & \colhead{Lookback Time (Gyr)$^a$} &
\colhead{DLAs} & \colhead{Detections} &
\colhead{Upper Limits} 
& \colhead{${\rm [}\langle{\rm Zn/H}_{\rm DLA}\rangle{\rm ]}$} 
}
\startdata
Full Sample  &  0.6922 $-$ 3.3901 & ~7.8 $-$ 14.7 & 34 & 19 & 15  & $-1.13 \pm 0.38$  \nl
& & & & & \nl
Subsample 1 &  0.50 $-$ 1.49 & ~6.3 $-$ 11.4 & 4  &  4 &  0  & $-0.98 \pm 0.33$ \nl
Subsample 2 &  1.50 $-$ 1.99 & 11.4 $-$ 12.7 & 8  &  6 &  2  & $-0.96 \pm 0.44$ \nl
Subsample 3 &  2.00 $-$ 2.49 & 12.7 $-$ 13.6 &12  &  6 &  6  & $-1.23 \pm 0.38$ \nl
Subsample 4 &  2.50 $-$ 2.99 & 13.6 $-$ 14.3 & 7  &  3 &  4  & $-1.11 \pm 0.27$ \nl
Subsample 5 &  3.00 $-$ 3.49 & 14.3 $-$ 14.8 & 3  &  0 &  3  & $\leq -1.39$ \nl
\enddata
\tablenotetext{a}{$H_0 = 50$~km~s$^{-1}$~Mpc$^{-1}$; $q_0 = 0.01$}
\end{deluxetable}
%\end{document}

\newpage 
\begin{figure}
\figurenum{1a}
\epsscale{1.1}
\plotone{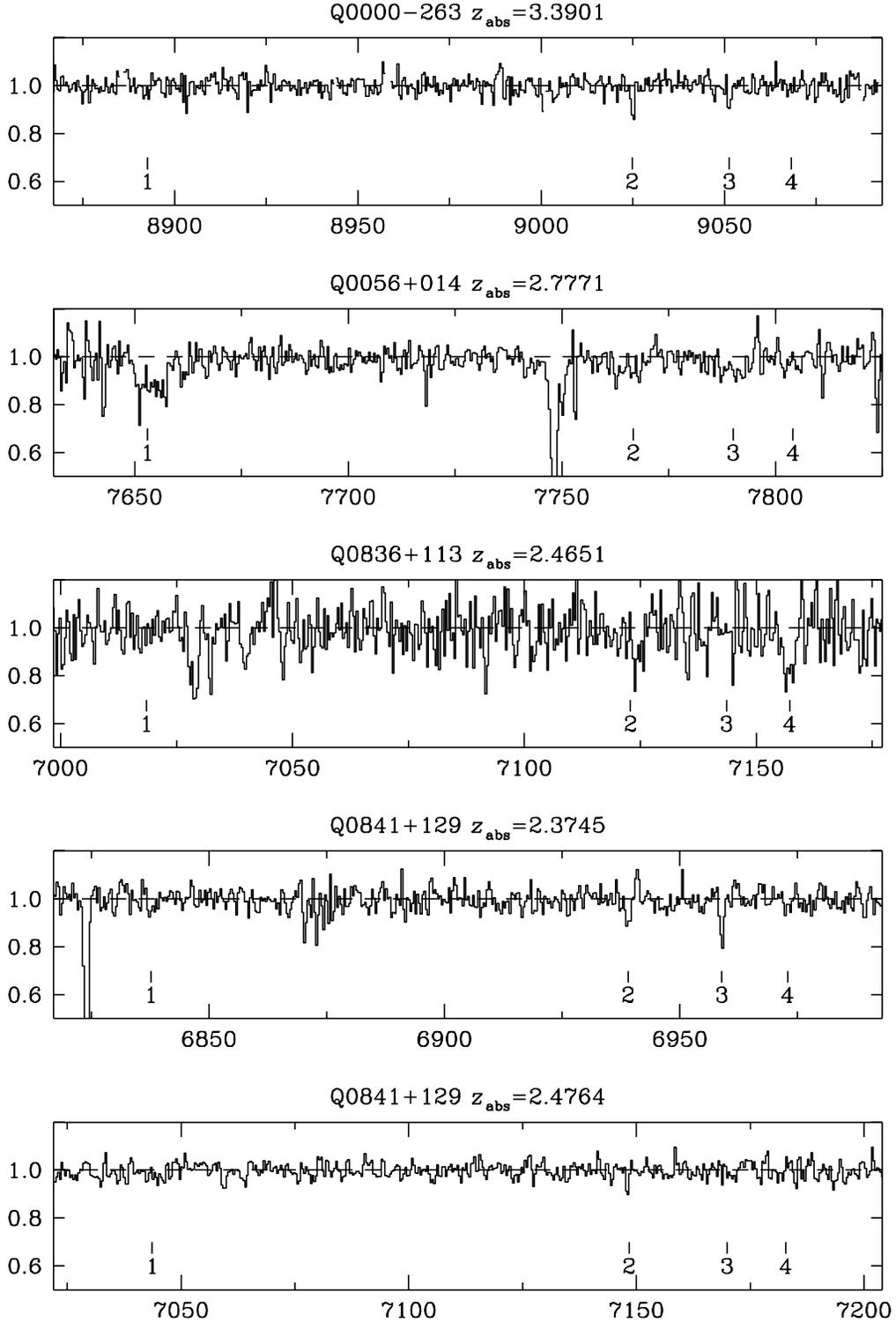}
\vspace{-3cm}
\caption{Portions of the QSO spectra observed in our survey of 
Zn~II and Cr~II lines in damped \lya\ systems. 
The $x$-axis is wavelength in \AA; the $y$-axis is residual 
intensity.
The vertical tick  marks  indicate
the expected positions of the absorption lines, 
whether they have been detected
or not. Line 1:~Zn~II~$\lambda 2025.483$; 
line 2:~Cr~II~$\lambda 2055.596$; 
line 3:~Cr~II~$\lambda   2061.575$~+~Zn~II~$\lambda  2062.005$  (blended);  
and  line 4:~Cr~II~$\lambda 2065.501$.  
The spectra have been normalised to the
underlying continua and are shown on an expanded vertical scale.
} 
\end{figure}
\newpage
\begin{figure}
\epsscale{1.1}
\figurenum{1b }
\plotone{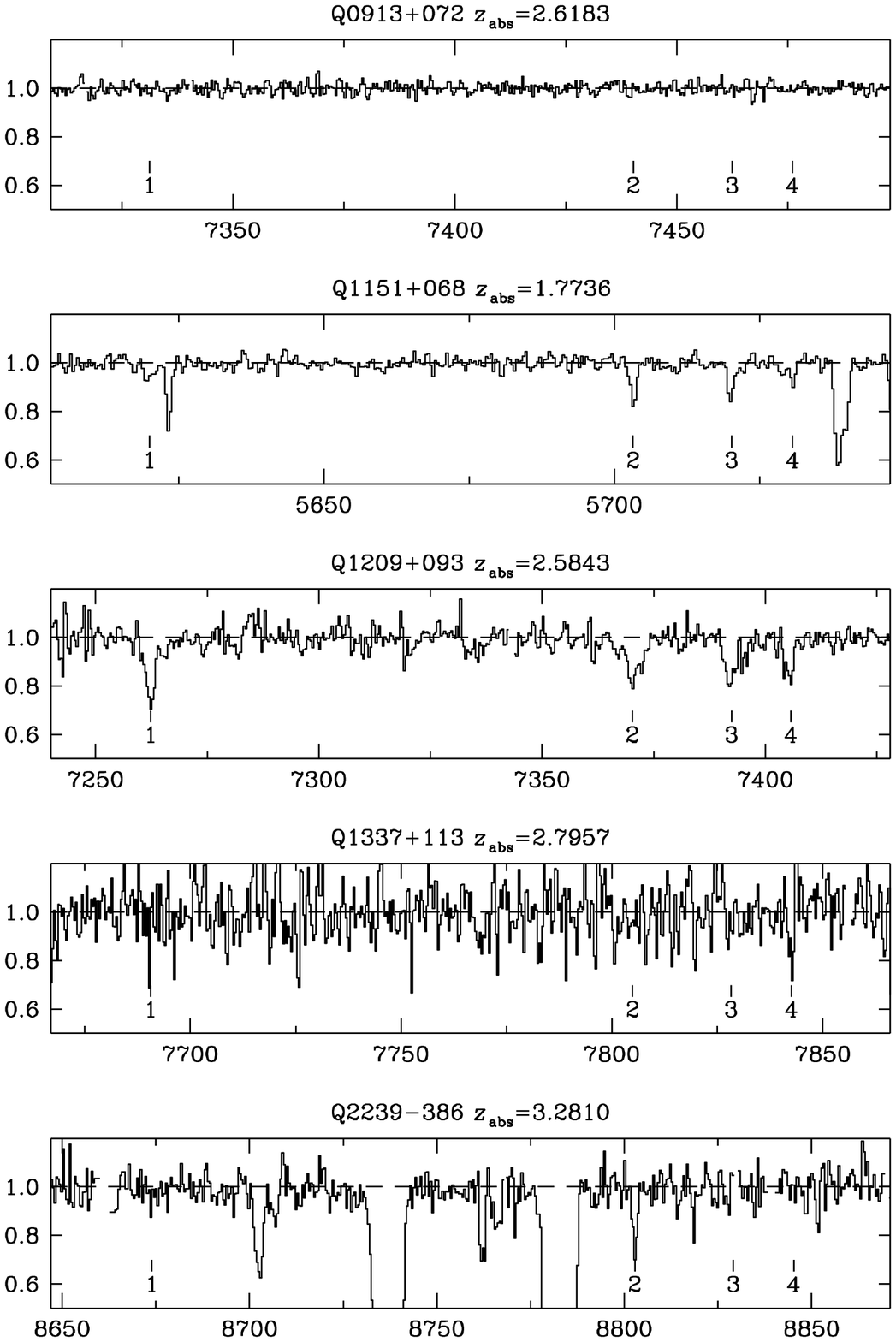}
\vspace{-1cm}
\caption{(continued)}
\end{figure}

\newpage 
\begin{figure}
\figurenum{2a}
\epsscale{1.1}
\plotone{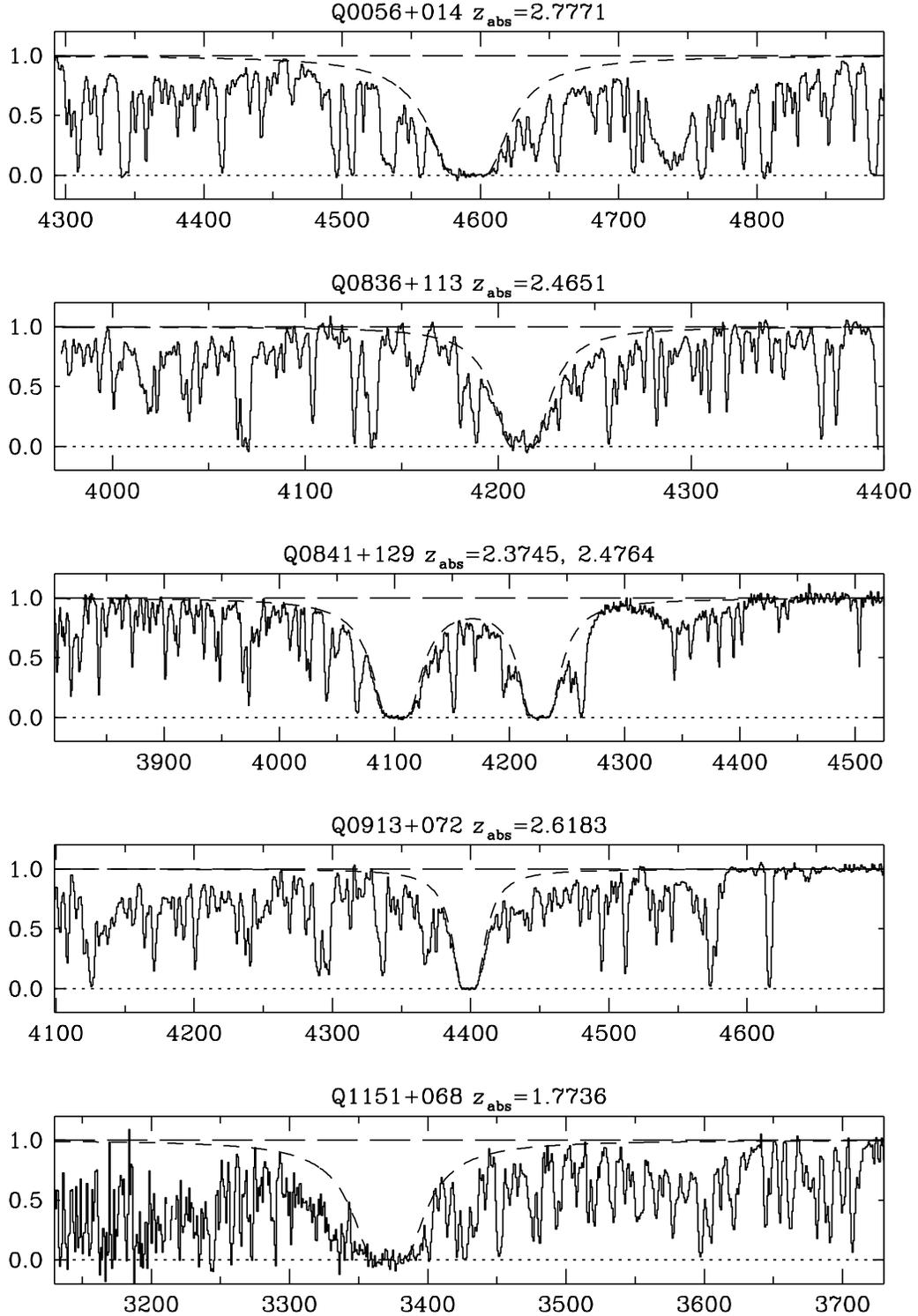}
\vspace{-2.5cm}
\caption{Normalised portions of the 
blue spectra of QSOs in our survey centred on 
the damped \lya\ absorption line. 
The $x$-axis is wavelength in \AA; the $y$-axis is residual
intensity.
In each panel the short-dash line shows the 
theoretical damping profile corresponding to 
the value of neutral hydrogen column density listed in 
column (3) of Table 4.
} 
\end{figure}
\newpage
\begin{figure}
\epsscale{1.1}
\figurenum{2b }
\plotone{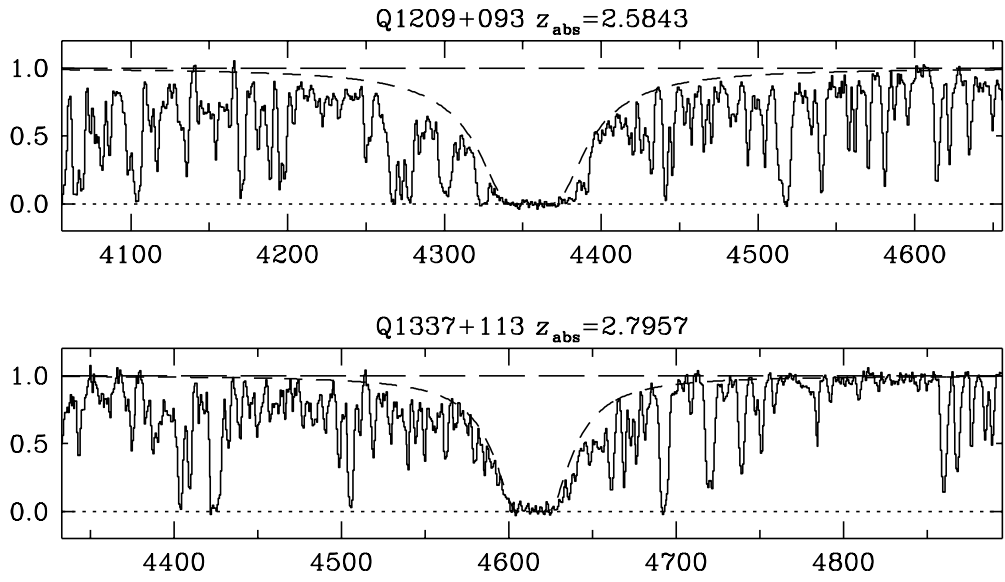}
\vspace{-3cm}
\caption{(continued)}
\end{figure}

\newpage 
\begin{figure}
\figurenum{3}
\epsscale{1.15}
\plotone{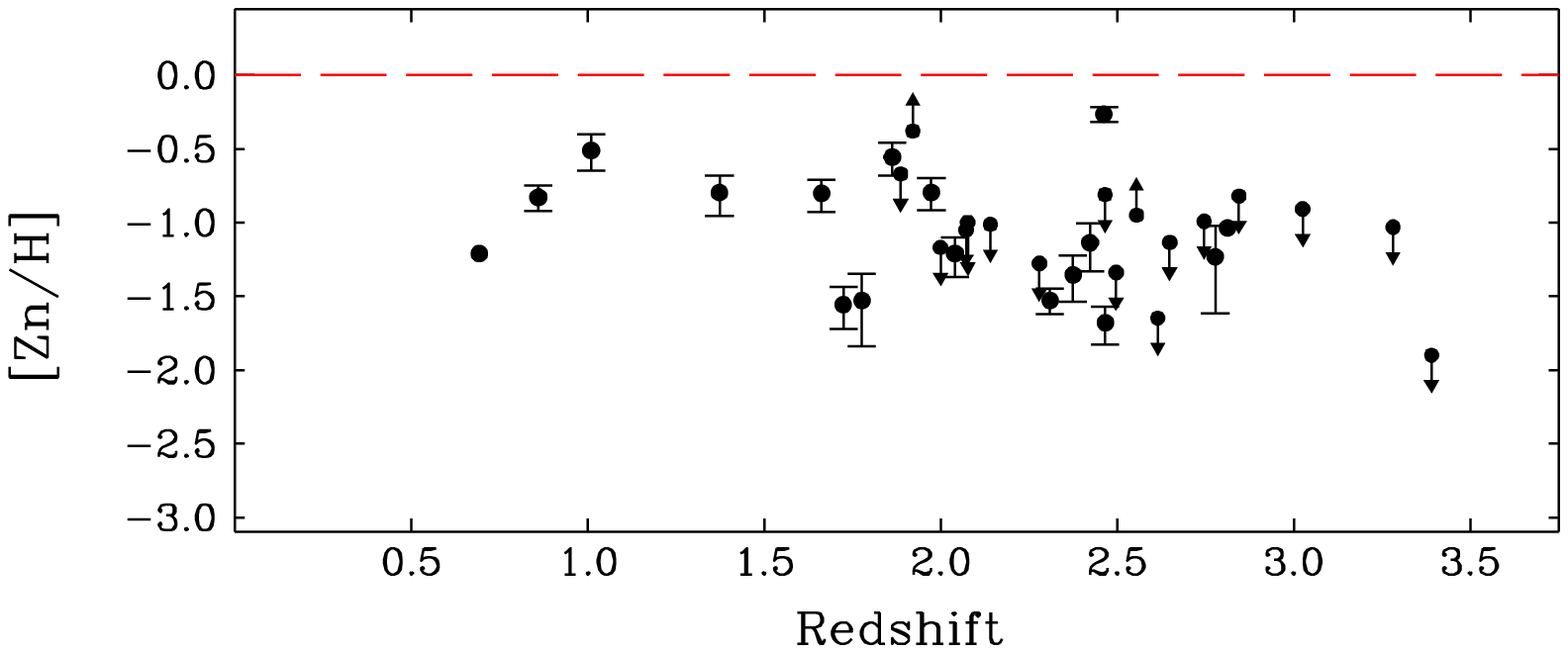}
\vspace{-10cm}
\caption{The abundance of Zn for the 34 damped  \lya\  systems
in the present survey plotted against redshift.  
Abundances are measured on a log scale relative to the solar value
shown by the broken line at [Zn/H] = 0.0\,.
Upper limits, corresponding to  non-detection  of  the  Zn~II 
lines, are indicated by downward-pointing
arrows. For two damped systems, indicated by dots with upward-pointing
arrows, we derive {\it lower} limits to the abundances, because the
absorption lines may be saturated. 
} 
\end{figure}

\newpage
\begin{figure}
\epsscale{1.1}
\figurenum{4}
\plotone{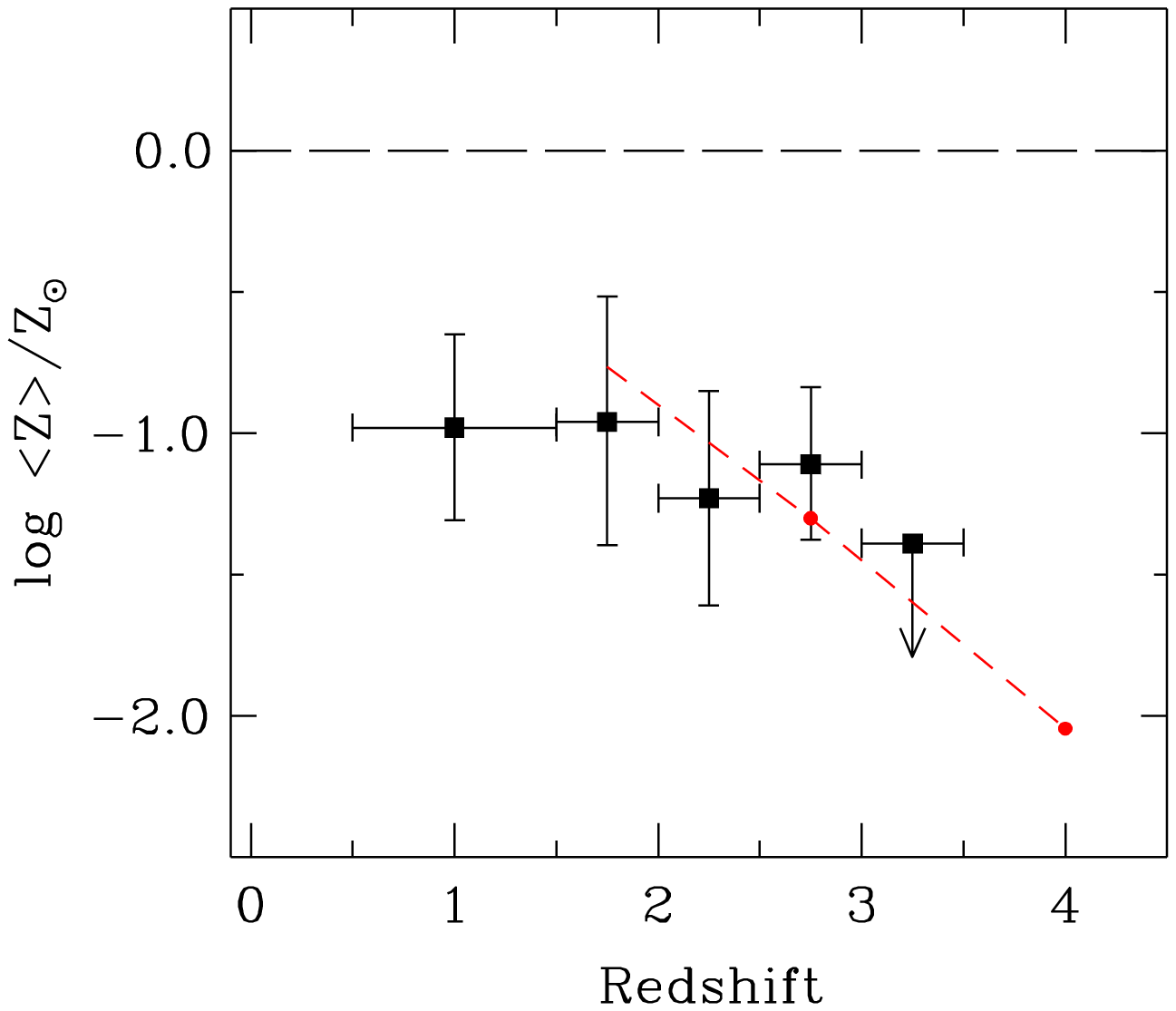}
\vspace{-6cm}
\caption{Cosmic chemical evolution as deduced from 
the metallicity of damped \lya\ systems (filled squares)
and from the metal ejection rate, $\dot{\rho_Z}$, 
of high redshift galaxies 
in the {\it Hubble Deep Field}
(dots).  
The filled squares are the values of 
${\rm [} \langle{\rm Zn/H}_{\rm DLA}\rangle {\rm ]}$
from the present survey;
the broken line shows the predictions of a simple integration of 
the redshift evolution of $\dot{\rho_Z}$ from Madau (1996) for
a $H_0 = 50$~km~s$^{-1}$~Mpc$^{-1}$ and $q_0 = 0.5$ cosmology.
For  $q_0 = 0.01$
the broken line would shift to lower values of
log~$\langle{\rm Z}\rangle/{\rm Z}_{\sun}$ by a factor of $\sim 2$\,. }
\end{figure}

\newpage
\begin{figure}
\epsscale{1.3}
\figurenum{5}
\plotone{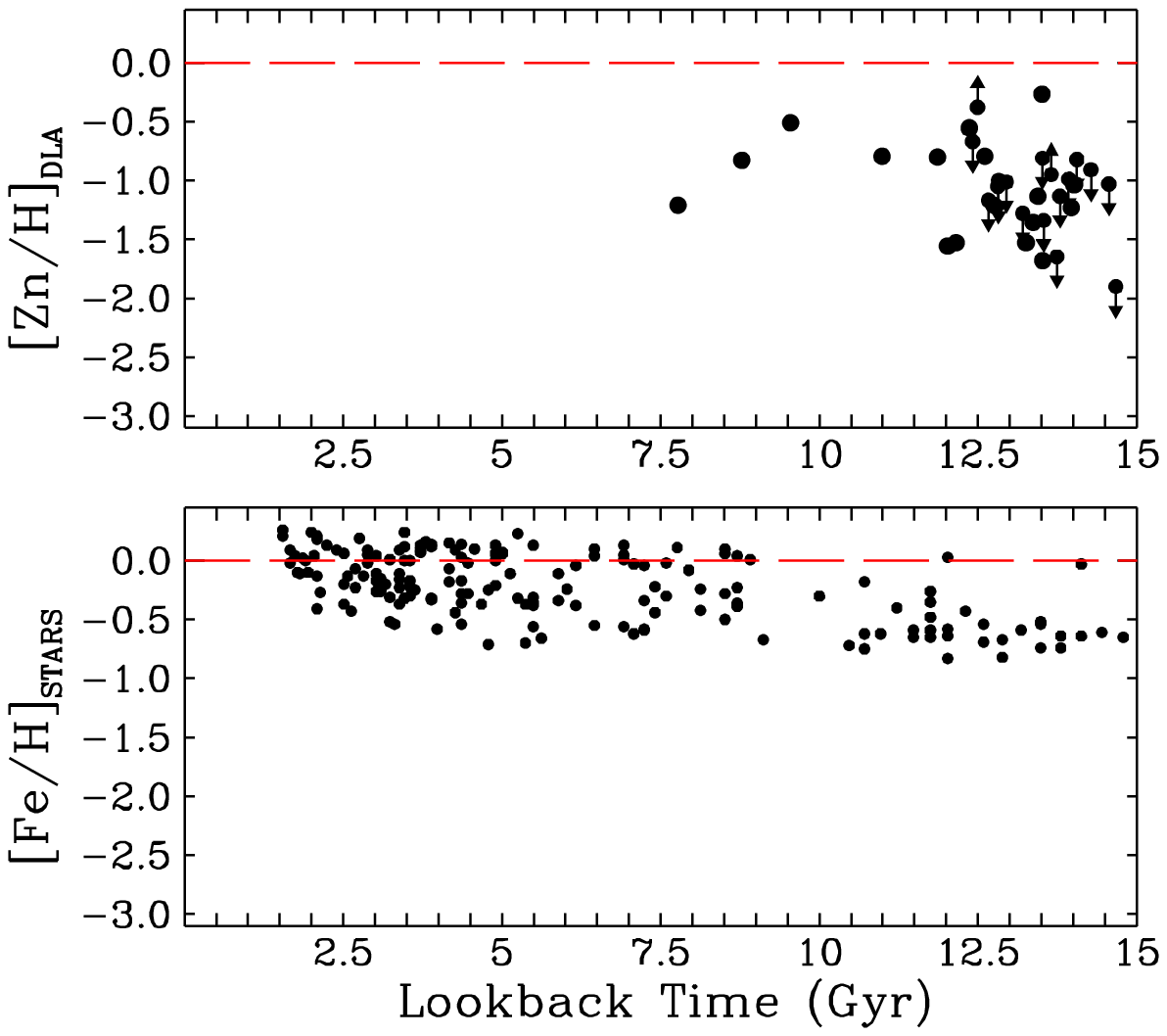}
\vspace{-10cm}
\caption{ 
{\it Top Panel:\/} Available measurements of  
metallicity in damped Lyman~$\alpha$ systems plotted as a function 
of lookback time for $H_{\rm 0} = 50$~km~s$^{-1}$~Mpc and 
$q_{\rm 0} = 0.01$\,.~
%The symbols have the same 
%meaning as in Figure 3.~ 
{\it Bottom Panel:\/} Metallicities of 182 F and G
dwarf stars in the Galactic disk 
with measured iron abundances and ages from the  
large-scale study by Edvardsson et al. (1993).}  
\end{figure}

\newpage
\begin{figure}
\epsscale{1.1}
\figurenum{6}
\plotone{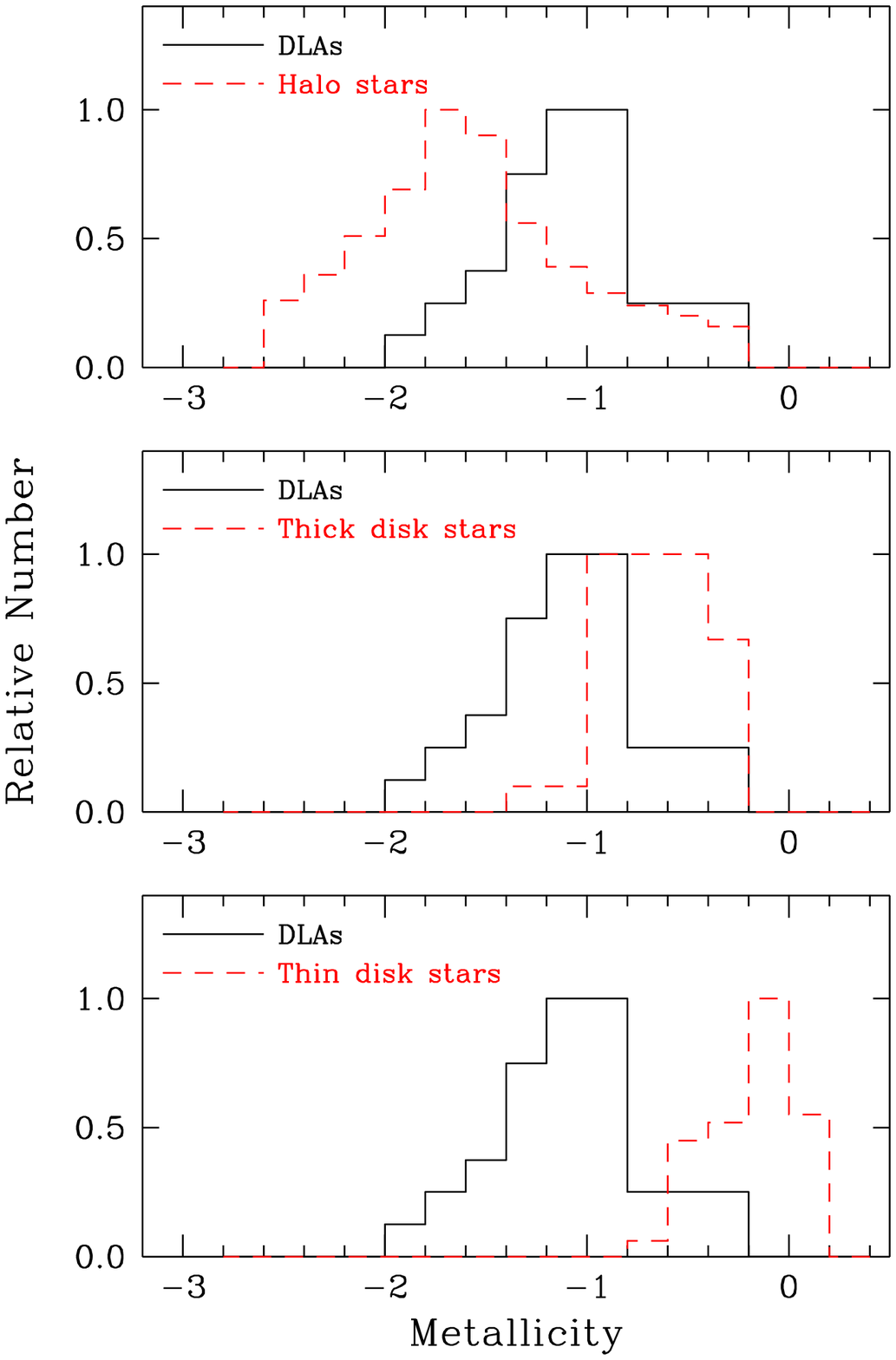}
\vspace{-3cm}
\caption{Metallicity distributions, normalised to unity,
of damped \lya\ systems and of stars belonging to the 
disk and halo populations in the Milky Way. 
See text for references to the 
sources of stellar data. 
Upper limits to [Zn/H]$_{\rm DLA}$
have been included as measurements in the histogram.
}
\end{figure}

\newpage
\begin{figure}
\epsscale{1.1}
\figurenum{7}
\plotone{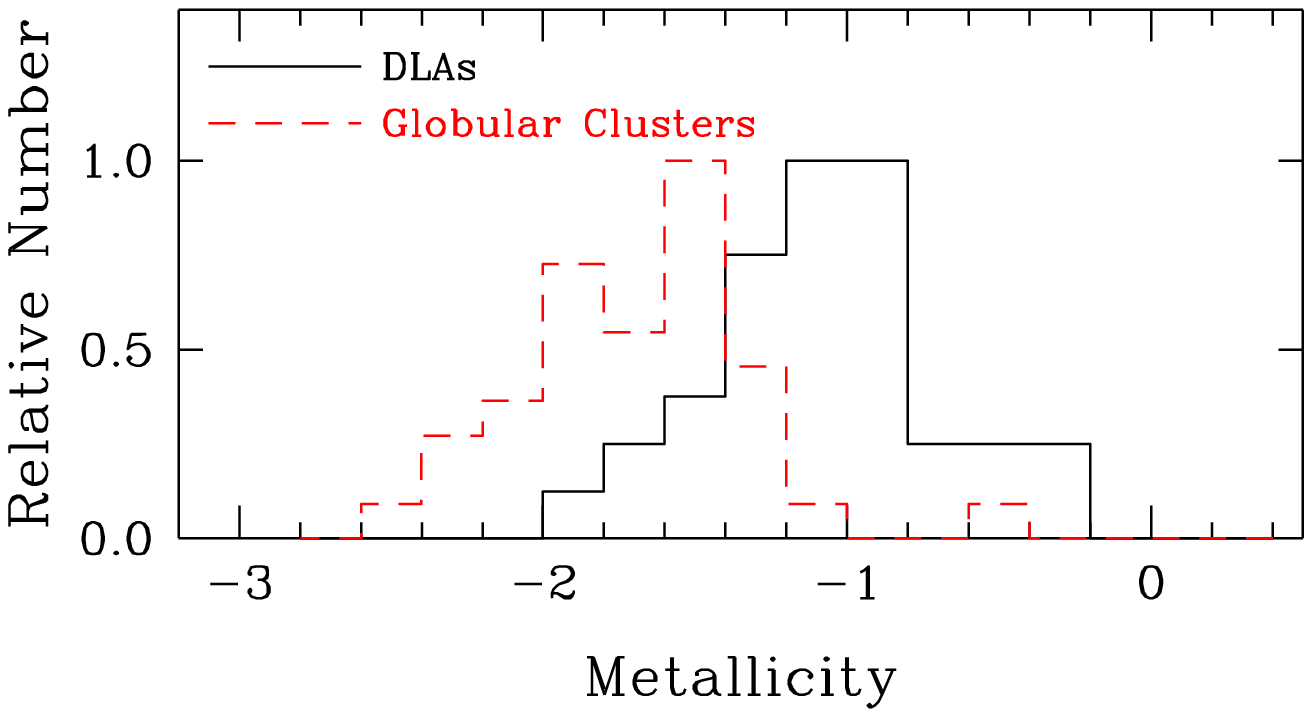}
\vspace{-6cm}
\caption{Metallicity distributions, normalised to unity,
of damped \lya\ systems and of 40 globular clusters in the sample
of Carney et al. (1996). 
Upper limits to [Zn/H]$_{\rm DLA}$
have been included as measurements in the histogram.
}
\end{figure}


\newpage

\begin{references}

\reference{} Anders, E., \& Grevesse, N. 1989, Geochim. Cosmochim. Acta, 
53, 197
\reference{} Baugh, C.M., Cole, S., Frenk, C.S., \& Lacey, C.G. 1997, 
ApJ, submitted (astro-ph/9703111)
\reference{} Beers, T.C., \& Sommer-Larsen, J. 1995, ApJS, 96, 175
\reference{} Bergeson, S.D., \& Lawler, J.E. 1993, ApJ, 408, 382
\reference{} Bowen, D.V., Blades, J.C., \& Pettini, M. 1996, ApJ, 472, L77
\reference{} Carney, B.W. 1996, PASP, 108, 900
\reference{} Carney, B.W., Laird, J.B., Latham, D.W., \& Aguilar, L.A. 
1996, AJ, 112, 668
\reference{} Chaffee, F.H., Foltz, C.B., Hewett, P.C., Francis, P.A.,
Weymann, R.J., Morris, S.L., Anderson, S.F., \& MacAlpine, G.M. 1991,
AJ, 102, 461
\reference{} \'{C}irkovi\'{c}, M.M., Lanzetta, K.M., Baldwin, J.,
Williger, G., Carswell, R.F., Potekhin, A.Y., \& Varshalovich, D.A. 1997, ApJ,
submitted
\reference{} Edvardsson, B., Andersen, J., Gustafsson, B., 
Lambert, D.L., Nissen, P.E., \& Tomkin, J. 1993,  A\&A, 275, 101
\reference{} Fall, S.M. 1996, in HST and the High Redshift Universe,
ed. N. Tanvir, A. Aragon-Salamanca, \& J.V. Wall (Singapore: World 
Scientific), in press. 
%\reference{} Fall, S.M., Charlot, S. \& Pei, Y.C. 1996, ApJ, 464, L43
\reference{} Fall, S.M., \& Pei, Y.C. 1993, ApJ, 402, 479
\reference{} Fan, X.M., \& Tytler, D. 1994, ApJS, 94, 17
\reference{} Freeman, K.C. 1991, in Dynamics of Disc Galaxies, 
ed. B. Sundelius (G$\ddot{o}$teborg University, G$\ddot{o}$teborg), 15 
\reference{} Friel, E.D. 1995, ARAA, 33, 381
\reference{} Ge, J., \& Bechtold, J. 1997, ApJ, in press (astro-ph/9701041)
\reference{} Gunn, J.E., \& Weinberg, D.H. 1995, 
in Wide Field Spectroscopy and the Distant 
Universe, ed. S.J. Maddox \& A. Aragon-Salamanca, (Singapore: World 
Scientific), 3
\reference{} Hazard, C. 1994, private communication
\reference{} Hazard, C., McMahon, R.G., \& Morton, D.C. 1987, MNRAS, 229, 371
\reference{} Hazard, C., Morton, D.C., McMahon, R.G., Sargent, W.L.W.,
\& Terlevich, R. 1986, MNRAS, 223, 87
\reference{} Hellsten, U., Dav$\acute{e}$, R., Hernquist, L.,
Weinberg, D.H., \& Katz, N. 1997, ApJ, in press (astro-ph/9701043) 
\reference{} Hunstead, R.W., Pettini, M., \& Fletcher, A.B. 1990, ApJ, 365, 23
\reference{} Ibata, R.A., \& Gilmore, G. 1995, MNRAS, 275, 605
\reference{} Laird, J.B., Rupen, M.P., Carney, B.W., \& Latham, D.W. 
1988, AJ, 96, 1908
\reference{} Lanzetta, K.M., \& Bowen, D.V. 1992, ApJ, 391, 48
\reference{} Lanzetta, K.M., Wolfe, A.M., \& Turnshek, D.A. 1995, ApJ, 440, 435
\reference{} Le Brun, V. Bergeron, J., Boisse, P., \& Deharveng, J.M. 
1997, A\&A, in press
\reference{} Levshakov, S.A., Chaffee, F.H., Foltz, C.B., \& Black, J.H. 
1992, A\&A, 262, 385
\reference{} Lu, L., Sargent, W.L.W., Barlow, T.A., Churchill, C.W., \& 
Vogt, S.S. 1996, ApJS, 107, 475
\reference{} Lu, L., Savage, B.D., Tripp, T.M., \& Meyer, D.M. 1995, 
ApJ, 447, 597
\reference{} Lu, L., \& Wolfe, A.M. 1994, AJ, 108, 44
\reference{} Lu, L., Wolfe, A.M., Turnshek, D.A., \& Lanzetta, K.M. 1993, 
ApJS, 84, 1
\reference{} Madau, P., 1996, in Star Formation
Near and Far, Proc. 7th Annual Astrophysics Conference in Maryland, 
ed. S.S. Holt \& G.L. Mundy (AIP: New York), in press (astro-ph/9612157)
\reference{} Madau, P., Ferguson, H.C., Dickinson, M., Giavalisco, M., 
Steidel, C.C., \& Fruchter, A. 1996, MNRAS, 283, 1388
\reference{} McGaugh, S.S. 1994, ApJ, 426, 135
\reference{} Meyer, D.M., Lanzetta, K.M., \& Wolfe, A.M. 1995, ApJ, 451, L13
\reference{} Meyer, D.M., Welty, D.E., \& York, D.G. 1989, ApJ, 343, L37
\reference{} Meyer, D.M., \& York, D.G. 1992, ApJ, 399, L121
\reference{} Molaro, P., D'Odorico, S., Fontana, A., Savaglio, S., \& 
Vladilo, G. 1996, A\&A, 308, 1
\reference{} Padoan, P., Jimenez, R., 
\& Antonuccio-Delogu, V. 1997, ApJ, in press
\reference{} Pei, Y.C., \& Fall, S.M. 1995, ApJ, 454, 69
\reference{} Pettini, M., Boksenberg, A., \& Hunstead, R.W. 1990, ApJ, 
348, 48 
\reference{} Pettini, M., \& Bowen, D.V. 1997, A\&A, submitted
\reference{} Pettini, M., King, D.L., Smith, L.J., \& Hunstead, R.W. 1995a, 
in QSO Absorption Lines, ed. G. Meylan (Berlin: Springer-Verlag), 71 
\reference{} Pettini, M., King, D.L., Smith, L.J., \& Hunstead, R.W. 1997,
ApJ, in press (April 1, 1997 issue) 
\reference{} Pettini, M., Lipman, K., \& Hunstead, R.W. 1995b, ApJ, 451, 
100
\reference{} Pettini, M., Smith, L.J., Hunstead, R.W., \& King, D.L. 1994,
ApJ, 426, 79
\reference{} Prochaska, J.X., \& Wolfe, A.M. 1996, ApJ, 470, 403 
\reference{} Prochaska, J.X., \& Wolfe, A.M. 1997a, ApJ, 474, 140 
\reference{} Prochaska, J.X., \& Wolfe, A.M. 1997b, in preparation
\reference{} Roth, K.C., \& Blades, J.C. 1995, ApJ, 445, L95
\reference{} Sargent, W.L.W., Boksenberg, A., \& Steidel, C.C. 1988, 
ApJS, 68, 539
\reference{} Savaglio, S., D'Odorico, S., \& Moller, P. 1994, A\&A, 281, 331
\reference{} Sembach, K.R., Steidel, C.C., Macke, R.J., \& Meyer, D.M. 
1995, ApJ, 445, L27
\reference{} Smette, A., Robertson, J.G., Shaver, P.A., Reimers, D., Wisotzki,
L., \& Kohler, T. 1995, A\&A Supp, 113, 199
\reference{} Smith, H.E., Cohen, R.D., \& Bradley S.E. 1986, ApJ, 310, 583
\reference{} Smith, L.J., Pettini, M., King, D.L., \& Hunstead, R.W. 1996,
in From Stars to Galaxies---the Impact of Stellar Physics on Galaxy
Evolution,  ed. C. Leitherer, U. Fritze-von Alvensleben \& J. Huchra,
Astr. Soc. Pacific Conf. Ser., 98, 559
\reference{} Steidel, C.C., Bowen, D.V., Blades, J.C., \& Dickinson, M.
1995a, ApJ, 440, L45 
\reference{} Steidel, C.C., Dickinson, M., Meyer, D.M., Adelberger, K.L., 
\& Sembach, K.R. 1997, ApJ, in press
\reference{} Steidel, C.C., Giavalisco, M., Pettini, M., Dickinson, M., \&
Adelberger, K.L. 1996, ApJ, 462, L17
\reference{} Steidel, C.C.,  Pettini, M., \& Hamilton. D. 1995b, AJ, 110, 
2519
\reference{} Steidel, C.C.,  Pettini, M., Dickinson, M., \& Persson, S.E. 
1994, AJ, 108, 2046
\reference{} Storrie-Lombardi, L.J., McMahon, R.G., \& Irwin, M.J. 1996a, 
MNRAS, 283, L79
\reference{} Storrie-Lombardi, L.J., McMahon, R.G., Irwin, M.J., \& 
Hazard, C. 1996b, ApJ, 468, 121 
\reference{} Taylor, K. 1995, in Wide Field Spectroscopy and the Distant 
Universe, ed. S.J. Maddox \& A. Aragon-Salamanca, (Singapore: World 
Scientific), 15
\reference{} Tripp, T.M., Lu, L., \& Savage, B.D. 1996, ApJS, 102, 239
\reference{} Turnshek, D.A., Wolfe, A.M., Lanzetta, K.M., Briggs, F.H., Cohen,
R.D., Foltz, C.B., Smith, H.E., \& Wilkes, B.J. 1989, ApJ, 344, 567
\reference{} Wolfe, A.M. 1995,
in QSO Absorption Lines, ed. G. Meylan (Berlin: Springer-Verlag), p.13
\reference{} Wolfe, A.M., Fan, X.-M., Tytler, D., Vogt, S.S., Keane, 
M.J., \& Lanzetta, K.M. 1994, ApJ, 435, L101 
\reference{} Wolfe, A.M., Lanzetta, K.M., Foltz, C.B., \& Chaffee, F.H. 1995, 
ApJ, 454, 698
\reference{} Wyse, R.F.G., \& Gilmore, G. 1995, AJ, 110, 2771
\reference{} Wyse, R.F.G., Gilmore, G., \& Franx, M. 1997, 
ARAA, 35, in press (astro-ph/9701223)
\reference{} York, D.G. 1988, in QSO Absorption Lines: Probing the Universe, 
ed. J.C. Blades, D.A. Turnshek, \& C.A. Norman 
(Cambridge: Cambridge Univ. Press), 227


\end{references}
\end{document}